\documentclass[pre,floatfix,superscriptaddress,showpacs]{revtex4}
\usepackage{amsmath}
\usepackage{graphicx}
\usepackage{epsfig}
\usepackage{latexsym}
\usepackage{amssymb}
\usepackage{ae}
\usepackage{aecompl}
\usepackage{hyperref}

\begin{document}
 \title{Phase transitions in nanosystems caused by interface
motion: The Ising bi-pyramid with competing surface fields}
 \author{A. Milchev}
 \affiliation{Institut f\"{u}r Physik, WA 331, Johannes Gutenberg
 Universit\"{a}t, D 55099 Mainz, Germany}
\affiliation{Institute for Physical Chemistry, Bulgarian Academy
of
 Sciences, 1113 Sofia, Bulgaria}
 \author{M. M\"{u}ller}
\affiliation{Institut f\"{u}r Physik, WA 331, Johannes Gutenberg
 Universit\"{a}t, D 55099 Mainz, Germany}
 \affiliation{Department of Physics, University of Wisconsin-Madison,
 1150 University Avenue, Madison, WI 53706-1390}
 \author{K. Binder}
 \affiliation{Institut f\"{u}r Physik, WA 331, Johannes Gutenberg
Universit\"{a}t, D 55099 Mainz, Germany} 


\begin{abstract} The phase behavior of a large but finite Ising
ferromagnet in the presence of competing surface magnetic fields
$\pm H_s$ is studied by Monte Carlo simulations and by 
phenomenological theory. Specifically, the geometry of a double
pyramid of height $2L$ is considered, such that the surface field is
positive on the four upper triangular surfaces of the bi-pyramid
and negative on the lower ones. It is shown that the total
spontaneous magnetization vanishes (for $L \rightarrow \infty$) at
the temperature $T_f(H)$, related to the ``filling transition'' of
a semi-infinite pyramid, which can be well below the critical
temperature of the bulk. The discontinuous vanishing of the
magnetization is accompanied by a susceptibility that diverges
with a Curie-Weiss power law, when the transition is approached
from either side. A Landau theory with size-dependent critical
amplitudes is proposed to explain these observations, and confirmed by
finite size scaling analysis of the simulation results. The
extension of these results to other nanosystems (gas-liquid
systems, binary mixtures, etc.) is briefly discussed.
\end{abstract}
\maketitle

\section{INTRODUCTION}
The current paradigm of attempting to develop various kinds of
nanoscopic devices requires careful consideration of the phase
behavior of nanosystems, since in nanoscopic geometries effects due
to external walls or other boundaries of the system can modify its
``bulk'' behavior substantially. Qualitatively new
kinds of phenomena may occur, that have not yet been studied for
macroscopic bulk systems.

We demonstrate a new kind of phase transition in the present
paper, which belongs to the class of interface
localization-delocalization phenomena, using the simple Ising
ferromagnet with nearest-neighbor exchange on a cubic
lattice as a generic example. Choosing a compact octahedral shape of the
system in the form of a bi-pyramid of height 2L, we assume that on
the upper surfaces of the pyramid ($0 < z \leq L$) a positive
surface magnetic field $+H_s$ acts, while on the lower surfaces
(with $-L\leq z <0$) the field is negative but of the same
absolute strength, so that no sign of the magnetization is overall
preferred. More generally, one might consider the case
with positive and negative fields of different strength;
their difference, however, could be effectively compensated by a suitably
chosen bulk field such that at low temperatures again a degeneracy
with respect to the sign of the spontaneous magnetization is
possible, similar to the case of ``capillary condensation''-type
phenomena in semi-infinite thin films
\cite{1,2,3,4,5,6,7,8,9,10,11}. In this case one can also expect an interesting interplay
between the wetting behavior of the semi-infinite system and the phase behavior
in confinement, a complication that is not considered in the present manuscript.

Such a system is then described (for $L 
\rightarrow \infty$) by an order parameter (the spontaneous magnetization 
of the Ising ferromagnet), which does \textit{not} remain non-zero up to the 
critical temperature $T_{cb}$ of the bulk three-dimensional model, but 
rather only up to a temperature $T_f(H_s)$, identical with the (critical)
temperature of the filling transition
\cite{12,13,14,15,16,17,18,19,20,21,22,23,24,25,26,27,28,29,30} in
a single semi-infinite pyramid. As will be discussed in this
paper, this new kind of phase transition \cite{31} in the limit $L
\rightarrow \infty$ can be either of first order or of second
order, depending on the value of the line tension, which describes
\cite{32,33,34,35,36,37} the free energy excess associated with
the contact line where the interface separating oppositely
oriented domains meets the free surface (or inert wall that
confines the system, respectively). Of course, as long as the
linear dimension L of the system is large but finite, finite-size
rounding of this phase transition needs to be considered, and
hence we shall present a tentative generalization of the theory of
finite size scaling \cite{38,39,40,41,42,43,44,45} to the present
situation here.

\begin{figure}[!t]
\epsfig{file=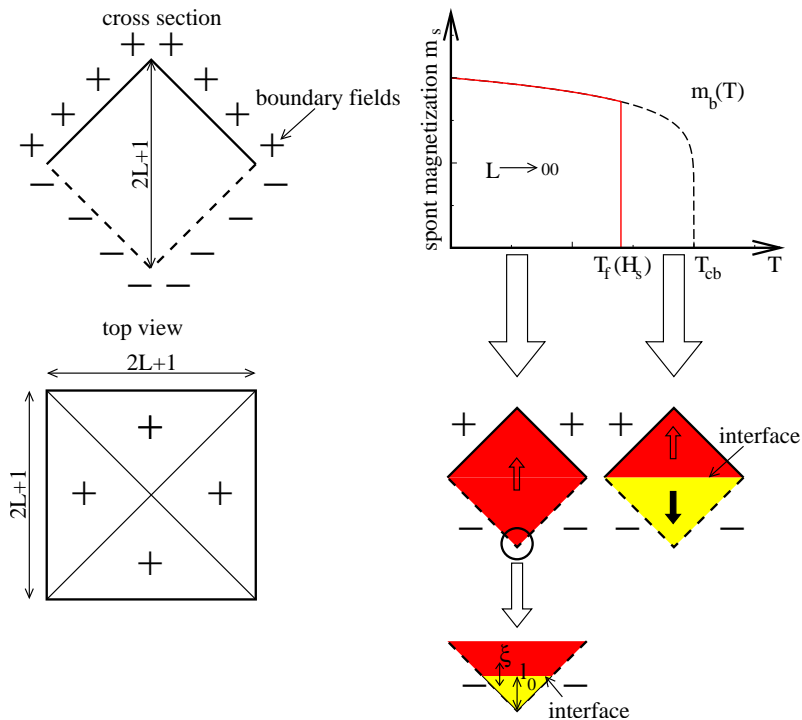,width=0.59\textwidth,clip=}
\caption{\label{fig1}Ising ferromagnet on a simple
cubic lattice whose surfaces form a bi-pyramid (left) and
resulting phase transition in the limit $L \rightarrow \infty$
(right), plotting the spontaneous magnetization $m_s(T)$ versus
temperature T. Signs $(+,-)$ along the cross section of the
bi-pyramid (left upper part) or on the triangular projections of
the surfaces in the top view (left lower part) refer to the
surface magnetic field, $\pm H_s$, that acts on the spins in the
surface planes only. Note that the basal plane of the bi-pyramid
is taken to be the (xy) plane of the simple cubic lattice, and
measuring lengths in units of the lattice spacing, each pyramid
takes L planes (with a single spin in the pyramid top), so the
total linear dimension from top to bottom of the bi-pyramid is
$2L+1$ (the extra lattice slice accounts for the basal plane common to both
pyramids). For $T < T_f(H_s)$, the interface between the domains
with negative ($\downarrow$) and positive ($\uparrow$)
magnetization is located close to one of the corners (e.g. the
bottom corner, as assumed in the figure, see the magnified view).
The local fluctuations of the interface extend over a correlation
range $\xi_\bot$, as indicated by the double arrow. As the
temperature is increased towards the filling transition
temperatures, $T\rightarrow T_f(H_s)$, the interface detaches from
the corner and moves towards the midplane of the bi-pyramid. For
$T >T_f(H_s)$ the magnetization $m_s$ then remains zero.}
\end{figure}
A qualitative explanation of this new transition is sketched in
Fig.~\ref{fig1}. It is assumed that the surface magnetic field
strength $H_s$ is small enough, so that for zero temperature the
ground state of the system has a uniform (positive or negative)
magnetization, in spite of the unfavorable energy cost due to the
surfaces (or walls, respectively) where the surface magnetic field
is oppositely oriented to the direction of the spontaneous
magnetization. As the temperature is raised, the interfacial free
energy $\sigma$ between oppositely oriented domains decreases faster than 
the excess free energy difference, $f_s(H_s,T)$, of a positively 
oriented domain between surfaces with $\pm H_s$. As is well known 
the interfacial free energy $\sigma$ vanishes at the bulk critical 
temperature $T_{cb}$ according to a power law $\sigma \propto 
(1-T/T_{cb})^{2\nu_b}$ with the correlation length critical 
exponent \cite{45a,45b} $\nu _b\approx 0.63$. At the temperature
$T_f(H_s)$, these surface free energies become equal, and hence
for $T>T_f(H_s)$ it is energetically favorable to have a state
with two oppositely magnetized domains, separated by an interface
located in the basal plane of the bi-pyramid (Fig.~\ref{fig1}). Of
course, in the actual calculation of $T_f(H_s)$ not only the
surface free energies $\sigma$ and $f_s(H_s,T)$ (per unit area) matter,
but one must also consider the fact that the four triangular
surfaces take a larger area (depending on the opening angle
$\alpha$ of the pyramids) than the area of the interface. In this
work, we consider explicitly only the case $\alpha = 45^{\circ}$,
so that the surfaces of the bi-pyramid meet at the basal plane at
an angle of $90^{\circ}$, but it is clear that the general
features of the phenomena described here do not depend on this
particular choice. In fact, we speculate that also the choice of
planar surfaces is an irrelevant detail, and similar behavior
could be observed for other geometries such as double-cones 
with different surface fields on the upper and lower portion.

The outline of this paper is as follows. In Sec. II, we recall the
basic facts about the filling transition in semi-infinite cones,
and develop a tentative phenomenological theory for describing the
transition explained in Fig.~\ref{fig1}. Sec. III describes our
Monte Carlo results and interprets them in terms of the
phenomenological description of Sec. II. Finally Sec. IV contains
our conclusions, and discusses briefly the extension to phase
transitions of other systems (gas-liquid systems, binary mixtures,
etc.) in related geometries.

\section{THEORETICAL BACKGROUND}
In this section two complementary phenomenological approaches to the phase transition
in a double-pyramid are developed. In subsection A we use the description
of the filling of a single cone \cite{21}. This yields the location of the 
filling transition in the limit $L \to \infty$ and we discuss modifications 
due to the double-pyramid geometry. This approach is expected to yield a 
good description if the magnetization is close to its saturation value, i.e.,
for $Lt \gg 1$, where $t$ denotes the reduced distance from the filling transition.
The role of fluctuations within this context is considered in subsection B.
Then, in subsection C, we develop a phenomenological Landau-type theory for the
case that the interface fluctuates around the basal plane. This approach is able
to describe the behavior in the ultimate vicinity of the transition, 
$L^2t \ll 1$, and the fluctuations above the transition.

\subsection{Phenomenological considerations in terms of surface
thermodynamics} Our phenomenological description assumes
that the theory of cone filling \cite{21} can be
directly applied to the filling of a semi-infinite
pyramid (i.e., we ignore the excess free energy at the edges of
the pyramid, where the contact lines of the interface with two
triangular pyramid surfaces meet). We compare
the bi-pyramid geometry to an equivalent situation of a
semi-infinite single pyramid, and consider the case when the
interface is located at a height $\ell _0$ above the bottom corner
\begin{figure}
\epsfig{file=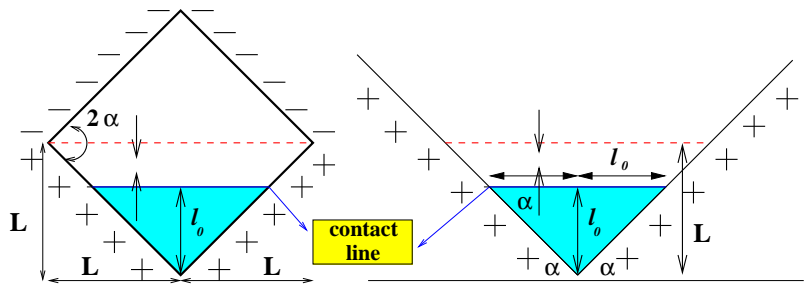,width=0.59\textwidth,clip=}
\caption{\label{fig2} Comparison of a bi-pyramid of
height $2L$ and basal plane of linear dimension $2L$ with a single
(infinitely large) pyramid, assuming the same positive surface
field $H_s$ at the corresponding surfaces, and considering the
situation that the interface between the domains with positive
$(\uparrow)$ and negative ($\downarrow$) magnetization is located
at the same distance $\ell _0$ from the bottom corner in both
cases. An opening angle $\alpha = 45^0$ is assumed for
simplicity.}
\end{figure}
(Fig.~\ref{fig2}). We write the free energy of the semi-infinite
pyramid, relative to a state with no interface, in terms of
surface and line free energies
\begin{equation}\label{eq1}
\Delta F_s=4\ell_0^2 \sigma + 8 \ell_0 \sigma _{\rm line} - 4 \sqrt{2}
\ell_0^2 f_s(H_s)
\end{equation}
In Eq.~\ref{eq1}, we have used the geometrical factors appropriate
for the opening angle $\alpha = 45^\circ$ (a generalization in
terms of other choices for the angle $\alpha$ is straightforward),
and we have also suppressed temperature arguments throughout, in
order to simplify the notation. Actually, rather than using the
temperature, $T$, as a control parameter as assumed in
Fig.~\ref{fig1}, we find it more convenient to use the strength of
the surface magnetic field $H_s$ instead. (In the plane of
variables $T,H_s$ the filling transition line is described by
the inverse function $H_{sc}(T)$ of the function $T_f(H_s)$. As
long as one crosses this line under a finite angle, it does not
matter whether $T$ or $H_s$ is used as a control variable).

Since we know that at the filling transition the interface can
move infinitely far apart from the lower corner, $\ell _0
\rightarrow \infty$, we must have $\Delta F=0$ for $H_s=H_{sc}$,
i.e.
\begin{equation}\label{eq2}
\sigma = \sqrt{2}f_s(H_{sc}).
\end{equation}
This result agrees with the macroscopic filling condition that the cone
fills if the contact angle on a planar substrate equals the cone angle, $\alpha$.
In the vicinity of $H_{sc}$ the variation of $f_s(H_s)$ with $H_s$
is linear, $f_s(H_s)=f_s(H_{sc})+(H_s-H_{sc})f_s'$. As a result,
near $H_s=H_{sc}$ Eq.~(\ref{eq1}) can be rewritten as
\begin{equation}\label{eq3}
\Delta F_s(\ell _0)=8\ell_0 \sigma _{\rm line} - 4 \sqrt{2} \ell_0^2
(H_s-H_{sc})f_s'
\end{equation}
Minimization of Eq.~(\ref{eq3}) with respect to $\ell_0$ readily
yields
\begin{equation}\label{eq4}
\ell _0 = \frac{\sigma_{\rm line}}{\sqrt{2}(H_s-H_{sc})f_s'}
\end{equation}
Since $f_s(H_s)$ is a monotonously increasing function of
$H_s,f_s'>0$ and hence the denominator of Eq.~(\ref{eq4}) is
{\em negative} in the considered region $H_s<H_{sc}$ (for $H_s > H_{sc}$
the pyramid is ``filled'', i.e., $\ell_0\equiv \infty$ on this other
side of the filling transition). Of course, only non-negative
solutions for $\ell_0$ are physically meaningful, and hence we
require that the line tension is negative, $\sigma_{\rm line}<0$. The
fact that critical cone filling can only occur for negative values
of the line tension has already been stressed by Parry et al.
\cite{21}. If $\sigma_{\rm line} \geq 0$, only a first order filling
transition is possible (i.e., at $H_s = H_{sc}$ the length $\ell_0$
jumps discontinuously from $\ell_0 =0$ to $\ell _0 = \infty$, in
our simplified treatment). Using Eq.~(\ref{eq4}) in
Eq.~(\ref{eq3}) yields
\begin{equation}\label{eq5}
\Delta F_s =2\sqrt{2} \sigma _{\rm line}^2
\frac{1}{(H_s-H_{sc})f_s'} .
\end{equation}
One should not worry about the fact that for $H_s \rightarrow
H_{sc}$ this free energy excess $\Delta F_s \rightarrow - \infty$,
because $\Delta F$ in Eqs.~(\ref{eq3}), (\ref{eq5}) is of order
unity only, rather than scaling with any power of the linear
dimension of the system. For the filling transition, the relevant
free energy scale is $4 \sigma L^2$, if for $H_s<H_{sc}$ we have
an interface of area $(2L)^2$ in the system. The free energy 
depression per unit area resulting from Eq.~\ref{eq5} is of order
$[(H_s-H_{sc})L^2]^{-1}$ and, hence, for $|H_s-H_{sc}|$ of order
$L^{-2}$ the divergence in Eq.~(\ref{eq5}) becomes problematic.
Taking the limit $L \rightarrow \infty$ first, and then letting 
$H_s \rightarrow H_{sc}$  obviously poses no problem: the free
energy per unit area stays $4\sqrt{2}f_s(H_s,T)$ for $H_s<H_{sc}$.

\begin{figure}
\epsfig{file=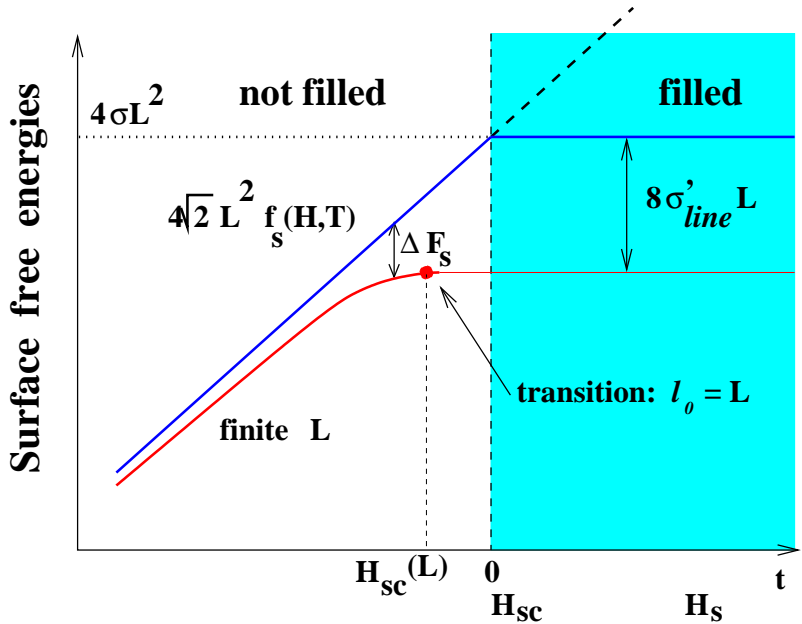,width=0.59\textwidth,clip=}
\caption{\label{fig3} Schematic plot of the surface
free energies of a large but finite bi-pyramid Ising system
versus $H_s$ (or $t=H_s/H_{sc}-1$, respectively). For
$L\rightarrow \infty$ the surface free energy is $4
\sqrt{2}L^2f_s(H,T)$ for $H_s<H_{sc}$, since the contribution due
to the interface at height $\ell _0$ above the lower corner is
negligible. For $H_s$ near $H_{sc}$ the variation of $f_s(H,T)$
with $H_s$ is linear. For $H_s>H_{sc}$ the surface free energy is
due to the interface in the basal plane of the bi-pyramid, $4
\sigma L^2$, independent of $H_s$, in the limit $L\rightarrow
\infty$. If large but finite linear dimensions L are considered,
the surface free energies are reduced because of line tension
contributions. For $H_s>H_{sc}(L)$, which is characterized by
$\ell _0 =L$ (dot in the figure), this reduction is $8 \sigma_
{\rm line}'L$ (in the vicinity of $H_{sc}$ the dependence of $\sigma
_{\rm line}' $ on $H_s$ can be neglected). For $H_s<H_{sc}(L)$, the
depression $\Delta F_s$ of the surface free energy relative to its
asymptotic expression for $L \rightarrow \infty$ gradually
grows with increasing $H_s$, reflecting the gradual increase
of $\ell _0$. Assuming that the gradual motion of $\ell _0
\rightarrow L$ can be described analogous to a
second-order-transition in a bulk system, the free energy is drawn
to meet the branch $4\sigma L^2-8 \sigma ' _{\rm line} L $ at
$H_{sc}(L)$ with horizontal slope.}
\end{figure}

We suggest now that in a large but finite bi-pyramid the behavior
of the surface free energies gets modified as schematically shown
in Fig.~\ref{fig3}. For $H>H_{sc}(L)$ the interface is located at
$\ell_0=L$, in the basal plane of the bi-pyramid, and the free energy 
is reduced by a line tension contribution, $f=4\sigma L^2+8\sigma 
'_{\rm line}L$. Note that, in general, the line tension of an interface in the basal 
plane of the bi-pyramid,  $\sigma _{\rm line}'$, where two planes (with 
surface fields $+H_s$ and $-H_s$) meet under an angle 2$\alpha$ 
(Fig.~\ref{fig2}, left part), can be expected to differ
from the line tension $\sigma_{\rm line}$ of an interface that meets a 
flat surface under an angle $\alpha$ (Fig.~\ref{fig2}, right part), 
with a surface field $+H_s$ on both sides of the interface. It is the 
latter quantity, however, which determines the scale
of the depression $\Delta F_s$ below the leading variation
$4\sqrt{2}L^2f_s(H_s)=4\sigma L^2+4\sqrt{2}L^2f_s'(H_s-H_{sc})$.
Since properties like the total magnetization of the 
bi-pyramid change continuously, when $\ell_0$ increases from small
values to $\ell_0=L$, it is assumed that the transition from the
state with broken symmetry  (the interface being located at
$\ell_0$ or $2L-\ell_0$, respectively) for $H_s<H_{sc}(L)$ to the
symmetric state where $\ell_0=L$ is a {\em second order} transition,
implying that the two branches of the surface free energy meet at
$H_s=H_{sc}(L)$ with a common tangent. In our notation (and in
Fig.~\ref{fig2}) we have allowed for a shift of $H_{sc}(L)$ due to
finite size from its asymptotic value $H_{sc}\equiv \lim_{L
\rightarrow \infty} H_{sc}(L)$. Of course, in reality we must
expect that for finite $L$ there is a rounding of the transition in
addition to the shift, and hence there does not exist any value
$H_{sc}(L)$ where singularities of the considered model
(Figs.~\ref{fig1} and \ref{fig2}) occur for finite $L$, but this
rounding of the transition can only be allowed for if
statistical fluctuations are taken into account.

If one could take the result for the free energy depression
through the formation of an interface, Eq.~(\ref{eq5}), literally,
the resulting behavior of the surface free energies would even be
slightly more complicated then conjectured in Fig.~\ref{fig3}. In
fact, if we consider the surface free energy for $H_s<H_{sc}(L)$
explicitly
\begin{equation}\label{eq6}
f(H_s)=4\sigma L^2+4\sqrt{2}L^2 (H_s-H_{sc})f_s'+\Delta F_s \;,
H_s <H_{sc}(L)
\end{equation}
and insert Eq.~(\ref{eq5}), we recognize that $f(H_s)$ exhibits a
maximum at
\begin{equation}\label{eq7}
H_s(L)_{max}-H_{sc}=\frac{\sigma_{\rm line}}{\sqrt{2}f'_s L},
\end{equation}
which corresponds to a value $\ell_0=L$. 
Using this result in Eq.~(\ref{eq6}), we readily find $f(H_s)=4\sigma L^2+8 \sigma_{\rm line}L$ at
the point marked by a dot  in Fig.~\ref{fig3}, as $H_s \to H_{sc}^-(L)$. On the other hand, approaching the 
transition from the other side ( $H_s \to H_{sc}^+(L)$ ), we have $f(H_s)=4\sigma L^2+8\sigma'_{\rm line}L$,
and there is a priori no reason to assume that $\sigma_{\rm line}=\sigma'_{\rm line}$ as noted above, because
of the physical distinction between the contact line on a plane and the contact line pinned to the edge where the
surface field changes sign (c.f.~Fig.~\ref{fig2}). So one would predict that in the surface free energy at $H_{sc}(L)$
there is a jump-singularity of order $L$. However, as will be discussed in the next section, thermal fluctuations are
expected to smooth out this singularity and, hence, $f_s(H_s)$ is a smooth non-singular function for all $L$.

\subsection{Interfacial fluctuations at filling transitions}
Here we recall results due to Parry et al. \cite{21} on the
filling of rotationally symmetric (infinite) cones. These authors
presented arguments that the dominating fluctuations of the
interface are the so-called ``breather modes'', i.e. the interface
moves \textit{uniformly} up or down. If we denote this fluctuating
height of the interface midpoint over the cone (or pyramid) corner
as $\ell$ and interpret $\ell_0$ of the previous subsection as its
average value, $\ell_0=\langle \ell \rangle$, the probability
distribution derived by Parry et al \cite{21} for the interface
position $\ell$ can be rewritten as
\begin{equation}\label{eq14}
P(\ell)\propto \exp\left[-\frac{1}{2\xi_\bot^2} (\ell - \ell_0)^2\right]\;,
\end{equation}
where the correlation length $\xi_\bot$ describing the interfacial
width due to these breather fluctuations diverges by a simple power law
\begin{equation}\label{eq15}
\xi_\bot\propto |t|^{-\nu_\bot}\;,\; \nu_\bot = \frac 1 2\;.
\end{equation}
Although Eqs.~(\ref{eq14}) and (\ref{eq15}) have been directly
obtained for axially symmetric cones only, Parry et al.
\cite{21} assert that they should hold as well for the inverted
pyramid-shaped geometry considered in the right part of
Fig.~\ref{fig2}.

It is interesting to apply Eqs.~(\ref{eq14}) and (\ref{eq15}) to such
a semi-infinite inverted pyramid for the case when $\ell_0$ has
reached the value $\ell_0=L$. Then Eqs.~(\ref{eq14}) and (\ref{eq15})
imply, using $\ell_0 \propto t^{-1}$,
\begin{equation}\label{eq16}
\langle \ell ^2\rangle - \langle \ell \rangle ^2 = \xi_\bot ^2
\propto |t|^{-1} \propto \ell _0 \propto L\;.
\end{equation}
This result also means, however, that the fluctuation in the
area of the interface is proportional to L as well.
Since these fluctuations of the interfacial free energy
are of the same order as the contribution of the line tension, a
theory based on balancing surface free energy differences with the
line tension alone, as sketched in the previous subsection, cannot
be expected to be quantitatively valid. Basically, one could argue,
what needs to be done is to average the free energy function of the previous section
with the Gaussian distribution
resulting from Eq.~(\ref{eq14}): the result will then be a
renormalized effective free energy varying smoothly with $H_s$, as
anticipated in Fig.~\ref{fig3}.

Eq.~(\ref{eq16}) yields a justification for the assumption that only
the uniform ``breather'' mode needs to be taken into account while
all the nonuniform interfacial fluctuations can be neglected. As
is well known, long wavelength nonuniform interfacial
fluctuations can be modeled as capillary waves
\cite{33,46,47,48,49,50}, and over a length scale L these
capillary waves cause a broadening of the interfacial profile
described by the following expression for the mean square width
$w^2$
\begin{equation}\label{eq17}
w^2=w^2_0 + \frac{k_BT}{4 \sigma} \ln \frac{L}{B}\;,
\end{equation}
where $w_0$ is a (hypothetical \cite{51}) intrinsic width and $B$ is
a short wavelength cutoff of the same order as $w_0$.
The logarithmic variation of $w^2$ with $L$ in Eq.~(\ref{eq17})
results from integrating the mean square amplitude $\langle
|h(\vec{q})|^2\rangle$ of the Fourier components $h(\vec{q})$ of
the deviation $h(\vec{x})=\ell(\vec{x})- \langle \ell \rangle$ of
the local height of the interface $\ell(\vec{x})$ from its average
value $\langle \ell \rangle$,
\begin{equation}\label{eq18}
\langle |h(\vec{q})|^2\rangle = \frac{k_BT}{\sigma q^2}
\end{equation}
over all wave numbers q in the interval $2\pi/L\leq q \leq
2\pi/B$.

The dominance of the uniform ``breather mode'' over the nonuniform
capillary waves is not unique to the problem of the filling
transition. Also in the problem of interface
localization-delocalization transitions in thin films\cite{5,52,53,54,55,56,57} 
of thickness $D$ a related anomalous size dependence
of interfacial widths was observed \cite{58}. Specifically, it was
found that for a fixed linear dimension L parallel to the
competing walls (in the ``soft mode'' phase \cite{52,53} where the
interface is unbound from the walls) the mean square fluctuation
of the interface scales even quadratically with $D$ \cite{58}
\begin{equation}\label{eq19}
w^2\propto \frac{k_BT \kappa}{\sigma} \frac{D^2}{L} \quad D
\rightarrow \infty \;, L \; \mbox{fixed}\;,
\end{equation}
corresponding to a fluctuation of the interface as a whole in the
direction normal to the interface over a finite fraction
of the film thickness. In Eq.~(\ref{eq19}), $\kappa^{-1}$ is a
length characterizing the exponential decay of the (short range)
repulsive effective potential acting on the interface from the
wall at $z=0$, $V(z)\propto \exp [-\kappa z] + \exp[-\kappa (D-z)]$, $z$
being the distance of the mean position of the interface from the
left wall. In the opposite limit, a linear variation of the mean
square fluctuation with D was found \cite{58,59}
\begin{equation}\label{eq20}
w^2=w^2_0 + \frac {k_BT\kappa D}{16 \sigma}+ {\rm const} \;,\; L
\rightarrow \infty, D \gg w_0\;.
\end{equation}
For a cubic geometry $D=L$ we note that Eqs.~(\ref{eq19}), (
\ref{eq20}) exhibit a smooth crossover characterized by
\begin{equation}\label{eq21}
w^2\propto \frac{k_BT}{\sigma} \kappa L\;,\; L \rightarrow
\infty\;,
\end{equation}
which is the same type of relation as found above in the context
of the filling transition, Eq.~(\ref{eq16}). Unfortunately, when
the interface average position coincides with the
bi-pyramid basal plane (Fig.~\ref{fig2}, left part), the effective
interface potential is not known to us, and hence we cannot
quantify the prefactor in the relation $w^2= \langle \ell ^2
\rangle - \langle \ell \rangle ^2 \propto L$ in this case.

\subsection{A phenomenological Landau-like theory for the phase
transition of the Ising bi-pyramid} 
As discussed in the previous sections, the phase transition sketched 
in Fig.~\ref{fig1} cannot
be understood solely from a macroscopic balance of surface and
line free energies, but interfacial fluctuations must be taken
into account, and the dominating fluctuation that needs to be
considered, is a uniform fluctuation of the position $\ell$ of the
interface around its average position $\ell_0$, in Fig.~\ref{fig2}
(left part). However, to a first approximation, $\ell_0$ is
related to the total magnetization per spin, $m$, by
\begin{equation}\label{eq22}
m/m_b=[1-(\ell_0/L)^3],
\end{equation}
where $m_b$ is the bulk magnetization of an (infinite) Ising
lattice at the same temperature. Here we use simple geometrical
relations, noting that the volume of the total bi-pyramid is
$8L^3/3$, the volume which has opposite orientation of the
magnetization is $4\ell_0^3/3$, and we have neglected any surface
contributions to the magnetization, assuming that the local
magnetization is everywhere $\pm m_b$ in the bi-pyramid, right up
to the surfaces and to the interface. We shall discuss corrections
to this approximation below.

Being interested in $|m|/m_b \ll 1$, it makes sense to transform
from $\ell_0 $ to $\tilde{\ell}_0=L-\ell _0$, i.e. we count the
interface distance from the basal plane rather than the lower
pyramid corner, to conclude that in this limit $m/m_b\approx 3
\tilde{\ell}_0/L$. Thus we conclude that $(\langle
m\rangle-m)/m_b\approx 3(\ell_0-\ell)/L$, and hence
Eq.~(\ref{eq14}) can be rewritten as (in the following we take
$m_b\equiv 1$ as the unit of the magnetization per spin)
\begin{equation}\label{eq23}
P(m)\approx \exp\left[-\frac{L^2}{18\xi_\bot^2}(m-\langle m
\rangle)^2\right].
\end{equation}
Remembering that $\xi_\bot \propto |t|^{-1/2}$
[Eq.~(\ref{eq15})] this is equivalent to
\begin{equation}\label{eq24}
P(m) \approx \exp \left[- {\rm const} L^2(-t)(m-\langle m \rangle )^2\right].
\end{equation}
Comparing this expression with the general fluctuation formula
\cite{42,60}
\begin{equation}\label{eq25}
P(m)\propto \exp \left[-\frac{V(m-\langle m
\rangle)^2}{2k_BT\chi}\right]\;,
\end{equation}
$V=8L^3/3$ being the volume of the system,
we immediately conclude that the susceptibility per spin $\chi$ at
the transition of Fig.~\ref{fig1} satisfies a Curie Weiss law for
$t<0$ (i.e., $H_s<H_{sc}$)
\begin{equation}\label{eq26}
\chi \propto L/|t|.
\end{equation}
Note that Eq.~(\ref{eq26}) holds only for $L^2|t|\gg1$, because only
then Eq.~(\ref{eq24}) is sharply peaked at $m \approx \langle m
\rangle$, and the Gaussian approximation for $P(m)$ holds. It is
also interesting to note that the ``critical amplitude'' \cite{61}
$\Gamma _-$ in the power law $\chi= \Gamma _- |t|^{-\gamma}$ is
proportional to L, i.e. divergent in the thermodynamic limit.
However, this fact is trivially understood, since the bulk
magnetic field creates a Zeeman energy $H\langle m \rangle
(8L^3/3)$, scaling with volume, while shifting the interface
between the oppositely oriented domains costs an energy
proportional to the interface area (of order $L^2$) only.

For $H_s >H_{sc}$ (i.e., $t>0$) we expect that the susceptibility per spin,
$\chi$, in the analogous relation where $\langle m \rangle \equiv
0$,
\begin{equation}\label{eq27}
P(m)\propto \exp \left[-\frac 1 2 V m^2/(k_BT\chi)\right],
\end{equation}
also scales proportional to L, because the above argument with the
Zeeman energy remains valid. Hence it is tempting to assume that
there holds a Curie-Weiss law analogous to Eq.~(\ref{eq26}) also
for $t>0$, and we suggest therefore that
\begin{equation}\label{eq28}
P(m)\propto \exp \left[-{\rm const}\; L^2|t|m^2 \right]\;, \; t>0,\; L^2|t|\gg 1\;,
\end{equation}
in analogy with Eq.~(\ref{eq24}). Now we also remember that for
$t<0$ there is a symmetry with respect to the sign of the
magnetization, so Eq.~(\ref{eq24}) for $H=0$ really needs to be
replaced by an expression that is symmetric with respect to the
sign of $\langle m \rangle = \pm m_0$,
\begin{eqnarray}\label{eq29}
P(m) &\propto& 
  \frac 1 2 \left\{\exp \left[- {\rm const}\; L^2(-t)(m-m_0)^2\right] + 
                   \exp \left[- {\rm const}\; L^2(-t)(m+m_0)^2\right] \right\}  \\ 
     & \approx & \frac 1 2  \exp \left[- {\rm const}\; L^2(-t)\frac{(m-m_0)^2(m+m_0)^2}{4m_0^2}\right]
       \qquad \mbox{for}\; |m| \approx m_0 \; \mbox{and} \; L^2|t|\gg 1. \nonumber 
\end{eqnarray}
It then is tempting to interpret Eqs.~(\ref{eq28}), (\ref{eq29})
as limiting cases of a Landau-type theory
\begin{equation}\label{eq30}
P(m)\propto \exp \left[-\frac 8 3 L^3 \frac{f_L(m)}{k_BT}\right],
\end{equation}
with an effective free energy density $f_L(m)$ \cite{62}
\begin{equation}\label{eq31}
f_L(m)=f_0+ \frac{1}{2L}rm^2+\frac 1 4 u_Lm^4-Hm,
\end{equation}
where we now have added the magnetic field. Moreover
Eq.~(\ref{eq31}) can be rewritten as
\begin{equation}\label{eq32}
f_L(m)=f_0+\frac {r}{4L}m^2_0-
\frac{r}{4Lm_0^2}(m-m_0)^2(m+m_0)^2 - Hm\;,
\end{equation}
and thus Eq.~(\ref{eq32}) immediately leads to Eqs.~(\ref{eq28}),
(\ref{eq29}) if $r=r_0t,\; r_0$ being a constant. Since
$m^2_0=-r/(Lu_L)$, we recover a mean-field exponent $\beta=1/2$
for the power law $m_0=\hat{B}|t|^\beta$, as expected for a Landau
type theory. Of course, such a mean-field relation for $m_0$ is
consistent with the jump of the magnetization expected in the
thermodynamic limit (Fig.~\ref{fig1}) only if the critical
amplitude $\hat{B}$ diverges as L tends to infinity.

The considerations discussed so far do not give any clue where the
basic nonlinearity responsible for deviations of $f_L(m)$ from
Gaussian behavior comes from. Disturbingly, the $t$-dependence of
$m_0$ (supposedly valid for $m_0/m_b \ll 1$) is inconsistent with
that which follows when one uses the expression $\ell_0 =
\hat{\ell}_0/|t|$ ($\hat{\ell}$ being a critical amplitude)
 from Eq.~(\ref{eq4}) in Eq.~(\ref{eq22}). The
approach of $m_0/m_b$ to its saturation value unity goes as 
$m_0/m_b-1\propto(L|t|)^{-3}$. Assuming that the term $u_Lm^4$
describes physical effects due to the existence of corners, which
take a fraction of $1/V\propto L^{-3}$ of the volume, a plausible
assumption is \cite{31} $u_L=u/L^3$ ($=u/L^d$ in $d$ dimensions).
As a consequence, one predicts
\begin{equation}\label{eq33}
m_0=L\sqrt{\frac{r_0}{u}}(-t)^{1/2}
\end{equation}
Of course, the condition $m_0/m_b \ll 1$ requires $L^2|t|\ll1$. As we
shall see below, in this regime all singular behavior is smeared
out due to finite size rounding and, hence, Eq.~(\ref{eq33}) is not
directly observable. The same problem occurs for the critical
isotherm, which follows from Eq.~(\ref{eq31}) for $r=0,\;, H\neq
0$,
\begin{equation}\label{eq34}
\left(\frac{\partial f_L(m)}{\partial m}\right)_{t,H} = u_Lm^3-H
= u(m/L)^3-H=0
\end{equation}
as
\begin{equation}\label{eq35}
m_{0,t=0}(H)=L(H/u)^{1/\delta}\;,\;\delta =3,
\end{equation}
i.e. the critical amplitude of the power law $m_0(H)|_{t=0}=
\hat{D}H^{1/\delta}$ scales again proportional to $L$. If we
generalize the problem to hyper-bi-pyramidic geometry in general
dimensionality $d$, the result $\Gamma_\pm \propto L$ remains
unchanged, while the other critical amplitudes become
$\hat{B}\propto L^{(d-1)/2},\;\hat{D}\propto L^{d/3}\quad$\cite{31}.
Finally we note that $f_L(m_0)=f_0+rm^2/(4L)=f_0-L(r_0t)^2/(4u)$
for $t < 0$, as expected from Fig.\ref{fig3}. For $|t|$ of order
$1/L^2$ the depression of $f_L(m)$ relative to $f_0$ is only of
order $L^{-3}$, i.e. negligible on the scale of Fig.~\ref{fig3}. 
Only for $|t| \sim 1/L$ a free energy of order $1/L$ is obtained.

The concept that the dominant statistical fluctuations are the
``breather modes'', i.e. fluctuations of the uniform
magnetization, fluctuations with a zero-dimensional phase space
\cite{21}, is reminiscent of the behavior of Ising-like systems in
high dimensionalities, $d>4$, which exhibit mean-field critical
behavior \cite{63} but nevertheless for a description of finite
size rounding these variations of the uniform magnetization need
to be taken into account. In brief, the statistical mechanics of
this latter problem is formulated \cite{43,44,45} in terms of a
partition function,
\begin{equation}\label{eq36}
Z=\int dm \exp\left[-L^df(m)/k_BT\right]\;,
\end{equation}
assuming a $d$-dimensional hypercubic lattice of linear dimension 
$L$ with periodic boundary conditions, where ($f_0$ is a constant)
\begin{equation}\label{eq37}
f(m)=f_0+ \frac 1 2rm^2+ \frac 1 4 um^4-Hm\;, \quad r=r't,\; t
=T/T_c-1\;.
\end{equation}
With the magnetization distribution,
\begin{equation}\label{eq38}
P_L(m)=Z^{-1}\exp \left[-L^df(m)/k_BT\right],
\end{equation}
its moments are then calculated as
\begin{equation}\label{eq39}
\langle m ^k \rangle = \int m^kP_L(m)dm
\end{equation}

In this problem, however, $u$ is a constant and does not depend on
L, nor does any other L-dependence appear in $f(m)$ as given in
Eq.~(\ref{eq37}). With a little algebra \cite{43,44,45} it is then
straightforward to show that for $H=0$ the moments $\langle m
^k\rangle$ (for k even, odd moments all vanish) scale as
\begin{equation}\label{eq40}
\langle m ^k \rangle =L^{-kd/4}\hat{M}_k(tL^{d/2})\;,
\end{equation}
$\hat{M}_k$ being scaling functions that can be explicitly derived
from Eqs.~(\ref{eq36})-(\ref{eq39}).

Here we follow exactly the same procedure, the only difference
being that Eq.~(\ref{eq37}) needs to be replaced by
Eq.~(\ref{eq31}), with $u_L=u/L^3$. Thus we obtain, writing
$\Psi=m/m_0$ and considering $H=0$ for simplicity
\begin{equation}\label{eq41}
P_L(\Psi)= \frac{\exp \left[ -\frac{m_0^2L^3}{3k_BT\chi '}(\Psi^2-1)^2\right]}
{ \int \limits ^{+\infty} _{-\infty} d\Psi \exp \left[-\frac {m_0^2L^3}{3k_BT\chi '}(\Psi ^2-1)^2\right]}
\end{equation}
It is seen that this distribution depends only on a single
parameter, namely,
\begin{equation}\label{eq42}
\frac{m^2_0L^3}{k_BT\chi
'}=L^2\frac{r_0}{u}t\frac{L^3}{L/(2r_0t)}=\frac 2 u (r_0tL^2)^2\;,
\end{equation}
and hence one finds that all moments are functions of this single
parameter as well
\begin{equation}\label{eq43}
\langle |\Psi|^k\rangle = f_k(tL^2)\;,
\end{equation}
where the scaling function $f_k$ is defined in terms of
Eq.~(\ref{eq41}) by simple integrals. Since
\begin{equation}\label{eq44}
\langle |m|^k\rangle = m_0^k \langle |\Psi|^k \rangle =
\left(\frac{r_0L^2t}{u}\right)^{k/2}\langle |\Psi|^k\rangle \equiv
\tilde{m}_k(tL^2)\;,
\end{equation}
also the scaling function $\tilde{m}_k$ is a function of the
scaling variable $tL^2$ again, there is no $L$-dependent prefactor,
unlike Eq.~(\ref{eq40}).

It is useful to consider the behavior right at $t=0$ separately,
since then (if also $H=0$) we have simply
\begin{equation}\label{eq45}
P_L(m)=\frac 1 Z \exp \left[\frac{-2um^4}{3k_BT}\right]\;,
\end{equation}
since the L-dependence of the volume $V=8L^3/3$ is canceled by
the L-dependence of $u_L,\; u_L=u/L^3$. From Eq.~(\ref{eq45}) it is
then obvious that all moments $\langle
|m|^k\rangle_{t=0}=\tilde{f}_k(0)$ are simple constants. If a
magnetic field is included, we similarly conclude
\begin{equation}\label{eq46}
\langle m^k\rangle = \tilde{m}_k(tL^2,HL^3)
\end{equation}
and in particular one finds that the zero-field susceptibility at
$t=0$ is finite and proportional to the volume, $\chi(t=0)\propto
L^3$. At this point we return to one - so far not really
justified - key assumption of the present treatment, namely
$u_L=u/L^3$. Equally well one could argue that the basic non-linearity, $u_Lm^4$,
of the effective free energy density is due to line tension
effects rather than caused by the presence of pyramid or cone corners, and hence
$u_L=u'/L^2$ would result. As a consequence, $m^2_0 =r_0|t|L/u'$, and
one would still predict a diverging amplitude of the order
parameter, $m_0=L^{1/2}\sqrt{r_0/u'}|t|^{1/2}$, and a finite size
scaling variable [Eq.~(\ref{eq42})] $m^2_0L^3/(k_BT\chi ')\propto
(tL^{3/2})^2$ instead of $(tL^2)^2$. The free energy then would
become $f_L(m)=f_0-(r_0t)^2/u'$, i.e. for $|t|L^{3/2}$ of order
unity it still would be of order $L^{-3}$, as expected, since the
regime of finite size rounding corresponds to total free energy
differences in the system of order $k_BT$. However, considering
the distribution $P_L(m)$ for $t=0$ we would obtain
$P_L(m)=Z^{-1}\exp \left[-\frac{2L^3}{3k_BT}u_Lm^4\right]=Z^{-1}
\exp\left[-\frac{2u'}{3k_BT}Lm^4\right]$, implying that for $t=0$ the moments
scale as $\langle |m|^k \rangle \propto L^{-k/4}$. As will be
demonstrated in Sec. III, such a behavior clearly contradicts
observation.

It is clear that the treatment presented so far is extremely
simplified, and one needs to discuss various corrections. For the
Landau theory of phase transitions in the bulk at high
dimensionalities [Eqs.~(\ref{eq36})-(\ref{eq40})] one knows that
nonuniform order parameter fluctuations are the leading source of
higher order correction terms to Eq.~(\ref{eq40}) \cite{64}. In
our problem, the analogous fluctuations to consider would be
nonuniform fluctuations of the interface. These fluctuations are
correlated in a volume $\xi _\bot ^d$ in $d$ dimensions, and the
motion of the interface back and forth in such a correlated volume
would cause a magnetization fluctuation of the order of
$m^2_b\xi_\bot^d$. Hence we conclude that a contribution $\Delta
\chi $ to the susceptibility results,
\begin{equation}\label{eq47}
k_BT\Delta \chi =L^d(\langle m ^2\rangle - \langle |m|\rangle ^2 )
\propto m_b^2\xi^d_\bot \propto |t|^{-d/2}
\end{equation}
We note that in $d=3$ this divergence has a stronger power than
the leading Curie-Weiss term, Eq.~(\ref{eq26}), but it has a
critical amplitude which is of order unity rather than of order $L$.
In the regime $|t|L^2\gg 1$ the asymptotic behavior predicted by
Eq.~(\ref{eq26}) should dominate, but for $|t|L^2$ of order unity
this correction may be non-negligible. Denoting the leading result
of Eq.~(\ref{eq26}) by $\chi _0$, we have in $d=3$
\begin{equation}\label{eq48}
\chi = \chi_0+\Delta \chi \propto L|t|^{-1} + {\rm const} \; |t|^{-3/2}
=L|t|^{-1}[1 + {\rm const}\; (|t|L^2)^{-1/2}]
\end{equation}
Thus in the region of the finite size rounding of the transition,
the nonuniform fluctuations yield corrections that are of the same
order as the corrections that would result from the scaling
function $\tilde{m}_k(tL^2)$ in Eq.~(\ref{eq44}). Hence one cannot
expect that moments $\langle |m|^k\rangle $ calculated from
Eq.~(\ref{eq41}) are quantitatively accurate, therefore, we have
not bothered to work them out in full detail. This result must be
expected, of course, because the result $\xi_\bot \propto |t|^{-1/2}$
means that for $|t|L^2$ of order unity $\xi_\bot$ is of order $L$,
the whole length scale of the Ising bi-pyramid.

It is also is of interest to consider the generalization of
Eq.~(\ref{eq48}) to arbitrary dimensionality $d$, which yields
\begin{equation}\label{eq49}
\chi = \chi_0+\Delta \chi = L|t|^{-1}[r_0 + {\rm const}\; /L|t|^{(d/2-1)}]
\end{equation}
This result shows that for ``hyper-bi-pyramids'' in $d>3$ $\Delta
\chi$ indeed becomes a correction, smaller than the terms
resulting from the scaling function in
Eqs.~(\ref{eq41})-(\ref{eq43}). Conversely, for $d<3$, the
corrections in the finite size scaling limit $|t|L^{d-1}$ of order
unity are proportional to $L^{d(3-d)/2}$ and hence larger than the
leading term of order $r_0$. This marginal role of $d=3$ may
indicate possible logarithmic corrections to finite size scaling.
For $d<3$ we expect a nontrivial description of finite size
effects with non-mean-field exponents. Monte Carlo studies of the
$d=2$ square geometry with competing edge fields \cite{65}
corroborate this conclusion.

In addition to the effects of nonuniform interfacial fluctuations,
there exist also corrections due to walls, edges and corners that
invalidate the simple relation between the interface height $\ell
_0$ and the magnetization, Eq.~(\ref{eq22}). E.g., from the regime
where in Fig.~\ref{fig2} a negatively oriented domain meets
positive surface fields, we expect a correction
$4\sqrt{2}(L^2-\ell_0^2)(\Delta m_s/m_0)/(8L^3/3)$ to
Eq.~\ref{eq22}, so that
\begin{equation}\label{eq50}
\frac{m}{m_b}=1-(\ell_0/L)^3-(3/\sqrt{2}) \frac{\Delta
m_s}{m_b}\left(\frac 1 L - \frac{\ell_0^2}{L^3}\right)\;,
\end{equation}
$\Delta m_s$ being a surface magnetization difference (per unit
surface area). Additional corrections (of order $L^{-2}$) may
result from the edges where the triangular surfaces of the pyramid meet.
Thus a phenomenological relation between $y=m/m_b$ and $x=\ell_0/L$ is
\begin{equation}\label{eq51}
y=1-A_0 x-A_1x^2-   A_2x^3,
\end{equation}
where $A_0,A_1$ and $A_2$ are phenomenological constants.

When $\ell_0$ is close to $L$, Eq.~(\ref{eq50}) can be simplified
as
\begin{equation}\label{eq52}
\frac{m}{m_b} =3 \left(\frac{L-\ell _0}{L}-\sqrt{2} \frac{\Delta
m_s}{m_b} \frac{L-\ell_0}{L^2}\right),
\end{equation}
which implies that in the linear relation between $(\langle
m\rangle - m)/m_b \approx 4 (\ell _0-\ell)/L$ used to justify
Eq.~(\ref{eq23}) a correction term (of relative order
$\sqrt{2}(\Delta m_s/m_b)/L$) enters, giving rise to the
replacement of the factor $L^2|t|$ in Eq.~(\ref{eq24}) by a linear
combination of terms $L^2|t|$ and $L|t|$, causing thus additional
corrections to finite size scaling.

\section{SIMULATION RESULTS}

\begin{figure}
\includegraphics{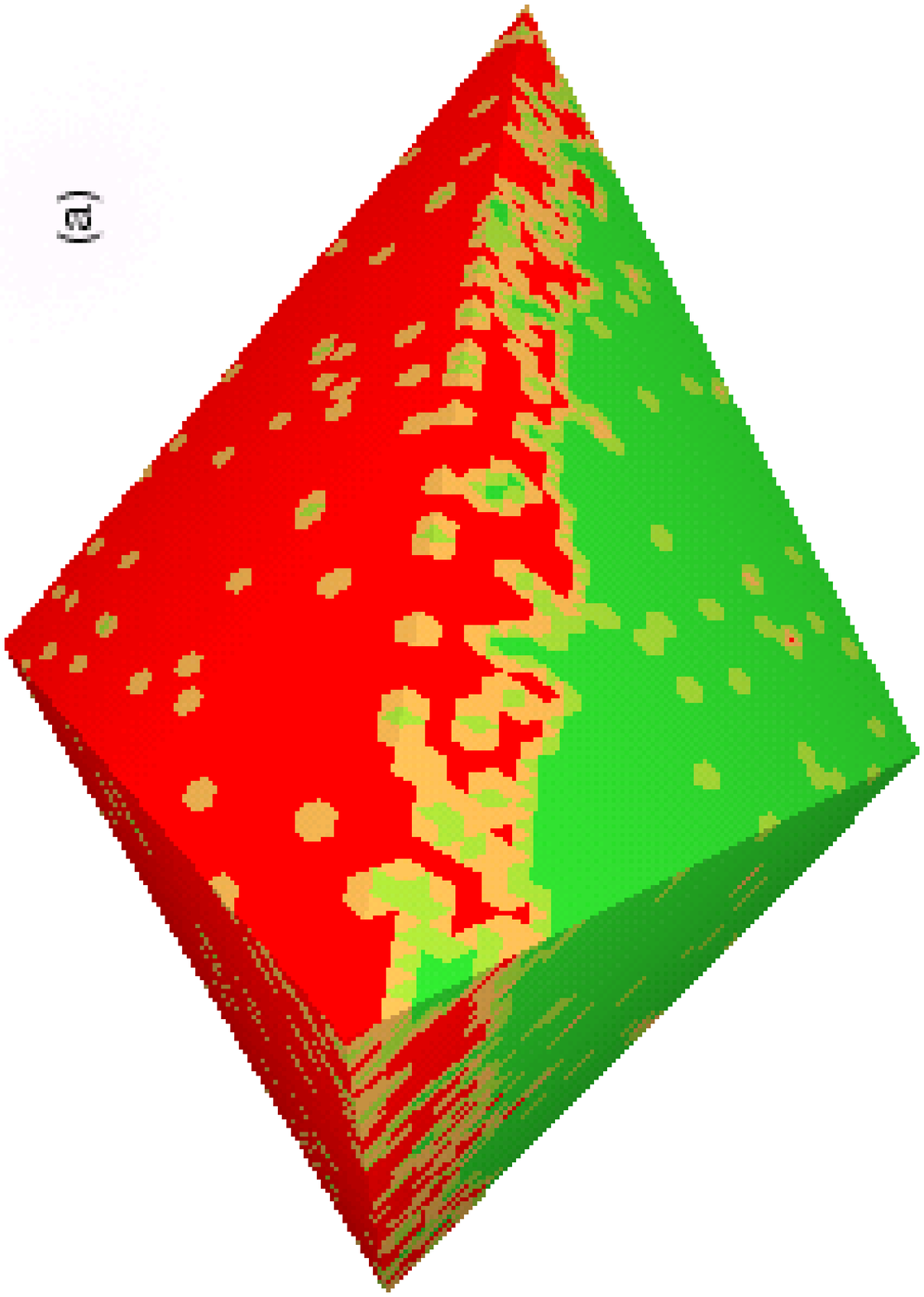}
\includegraphics{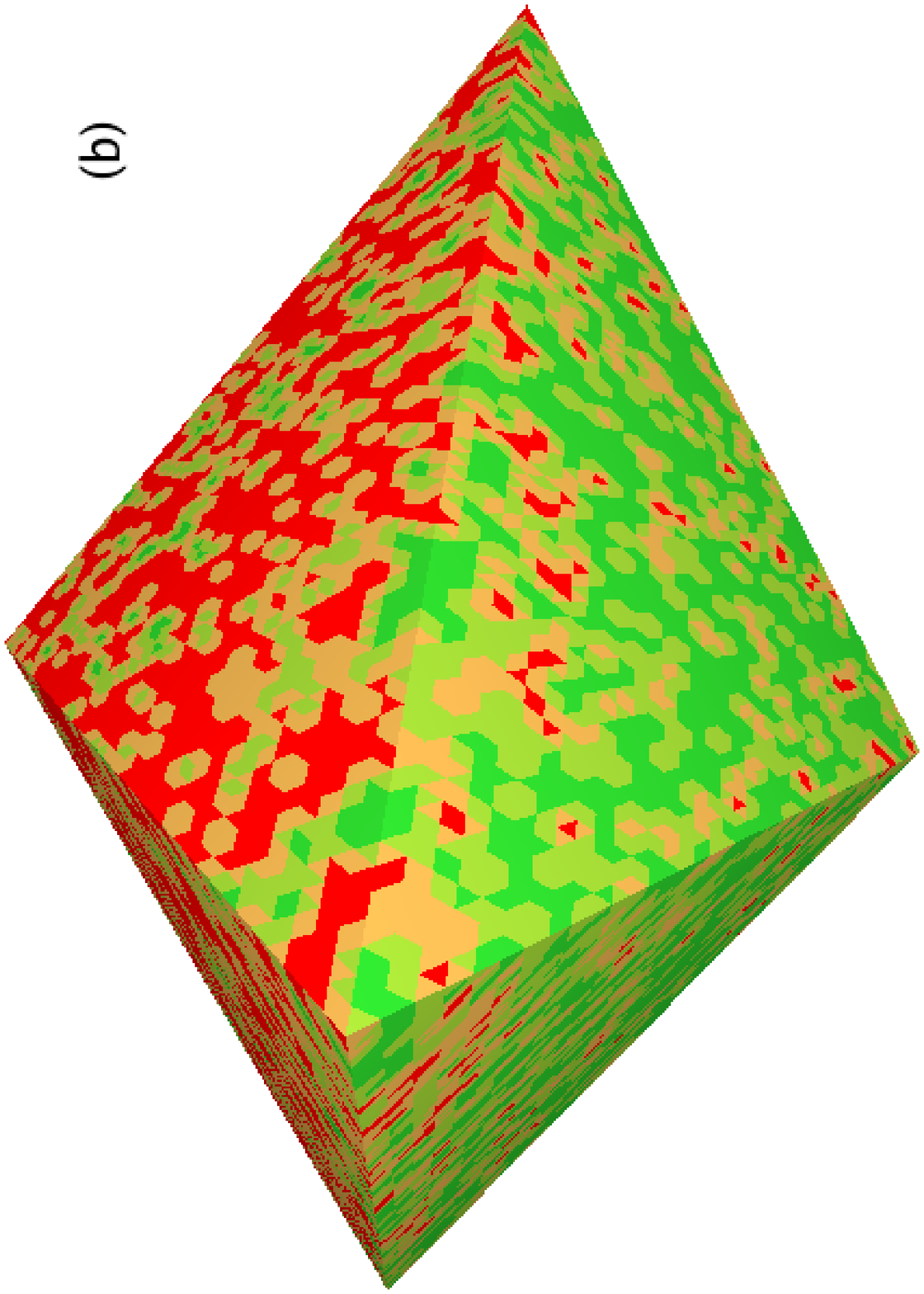}
\vskip 7.3cm
\caption{\label{fig5} Snapshot pictures of the state
of Ising bi-pyramids with $L=40$ and two surface fields, (a)
$H_s/J=1.00,J/k_BT=0.45$, and (b) $H_S/J=0.77, J/k_BT=0.25$.
The local magnetization $m_i$ at the pyramid surfaces is coarse-grained 
over each triangle of closest neighboring spins, thus taking on the values
$m_i = -1, -1/3, 1/3, 1$ with $i$ being the considered lattice site 
of the respective surface plane.} %In the following we choose units 
%$J \equiv 1$ and $k_B \equiv 1$.}
\end{figure}

Qualitative evidence for the actual occurrence of the transition
sketched in Fig.~\ref{fig1} is presented in Fig.~\ref{fig5},
showing two snapshots of the Ising bi-pyramid in the two
``phases'' caused by different values of the surface field. Note
that we use an Ising nearest neighbor Hamiltonian
\begin{equation}\label{eq53}
{\mathcal{H}}=-J \sum \limits _{\langle i,j \rangle}^{\rm bulk} S_iS_j
- J_s\sum \limits _{\langle i,j \rangle} ^{\rm surfaces}S_iS_j -
H_s\sum \limits _i ^{\rm upper\;surfaces}S_i + H_s \sum \limits _i ^{\rm lower\;
surfaces} S_i - H \sum \limits _i ^{\rm all\; spins} S_i \;,
\end{equation}
where the exchange constant is weakened if both spins i, j are in
a surface plane, $J_s = J/2$. One can see that for $J/k_BT=0.45$,
$H_s/J=1.00$ the magnetization is still predominantly negative, as
anticipated in the schematic drawing of Fig.~\ref{fig2} (left
part). In the negative domain only small clusters of positively
oriented spins occur, and vice versa. For $J/k_BT=0.25$ and
$H_s/J=0.77$, however, there is no majority of either positive or
negative spins, as far as one can tell this from viewing the
pyramid surfaces.

\begin{figure}
\includegraphics{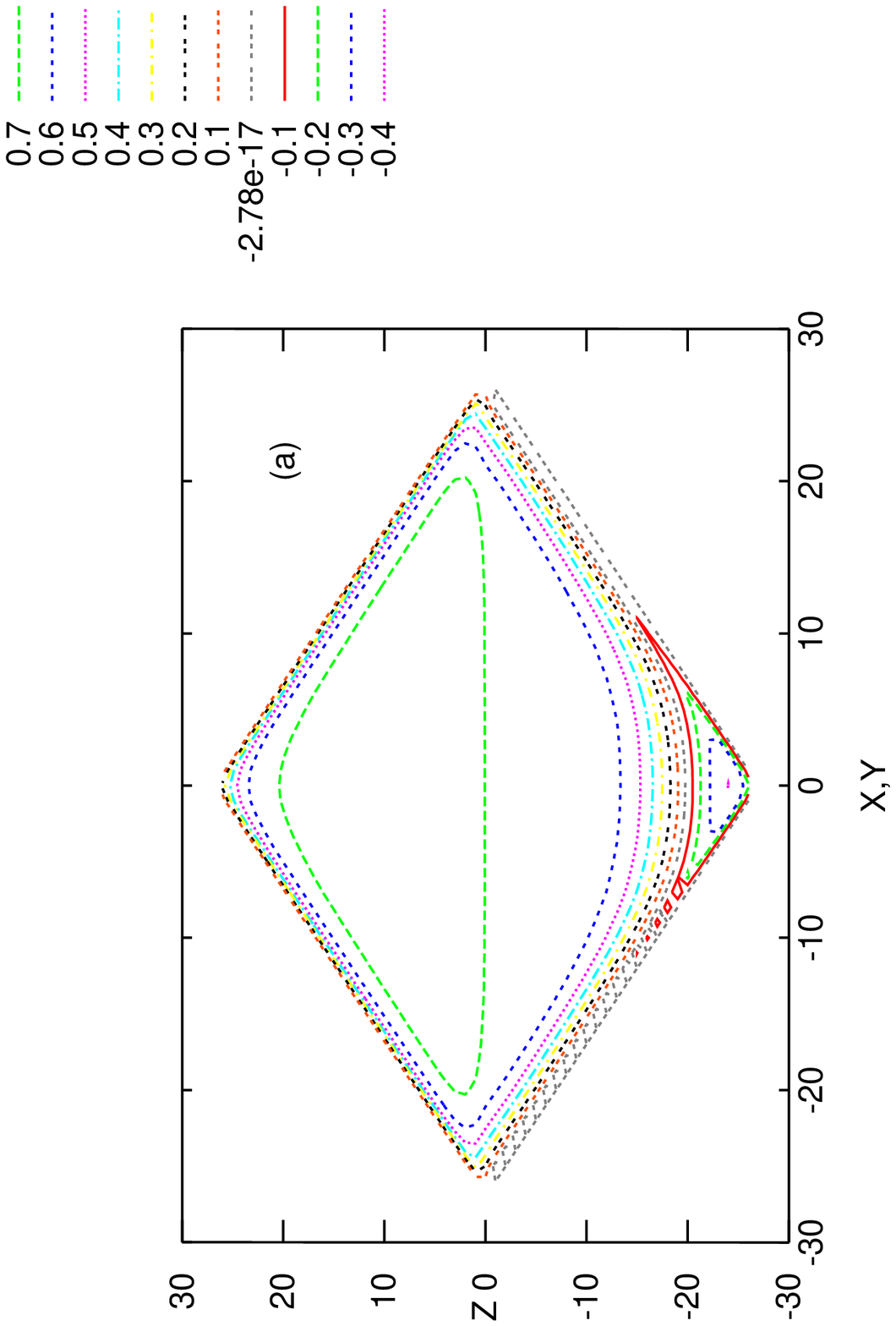}
\includegraphics{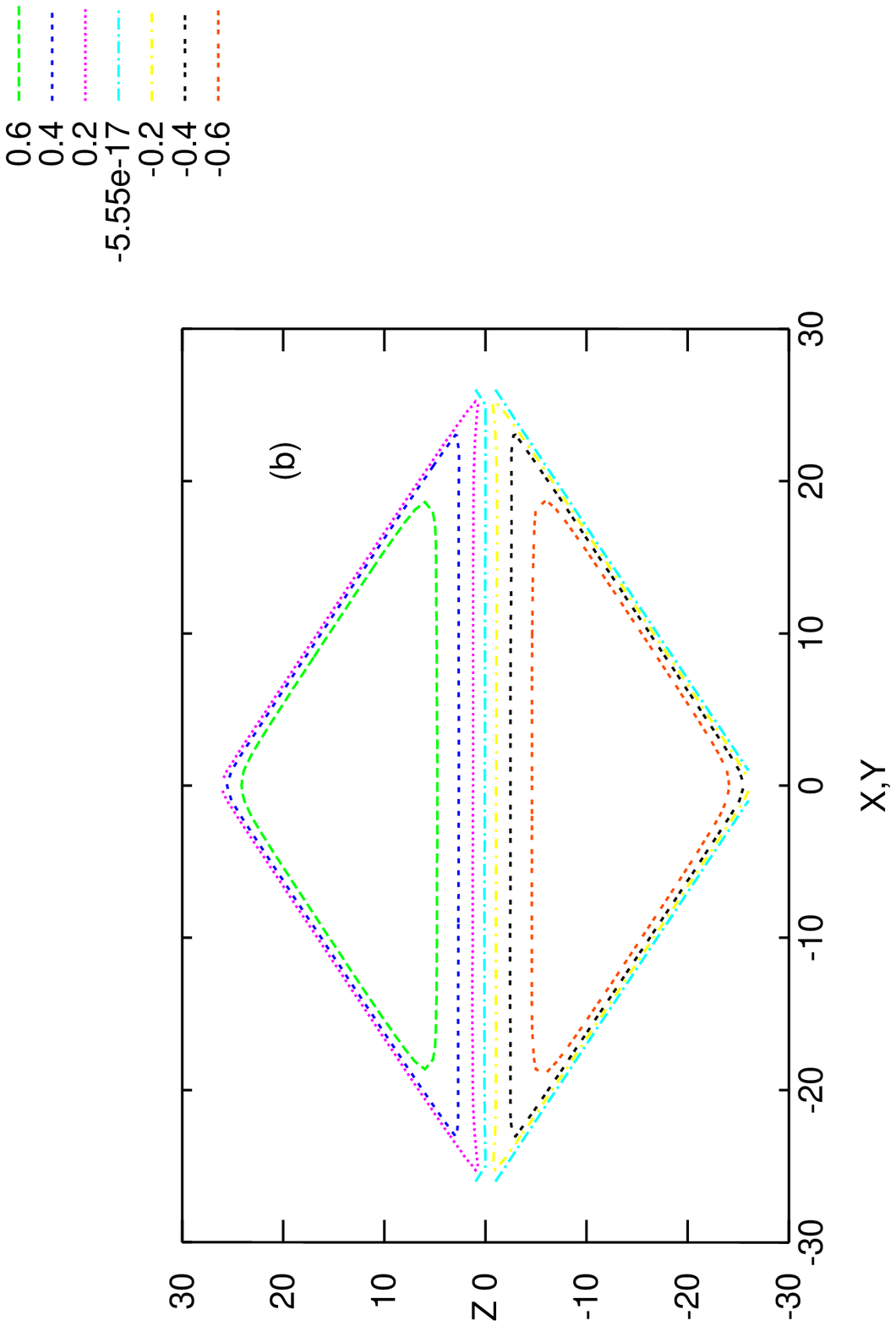}
\vskip 6.5cm
\caption{\label{fig6} Contour plots presenting
curves of constant magnetization (as shown in the key of each
figure) as a function of position (z and x or y, respectively,
choosing the coordinate origin in the center of mass of the
bi-pyramid) for $L=26,\; k_BT/J=4, \; J_s/J=1/2$, and $H_s=0.6$ (a) and 0.8 (b).}
\end{figure}

A more quantitative characterization of the transition is provided
by contour diagrams (Fig.~\ref{fig6}). Fig.~\ref{fig6} shows that
the schematic view of an interface at a height $\ell_0$ over the
pyramid corner (Figs.~\ref{fig1} and \ref{fig2}) is a very crude
over-simplification of reality: rather the interface is fairly
broad, spread out over a thickness of many lattice spacings. Moreover,
the interface is strongly bent and not at all horizontal. One
can also see that the interface is not hitting the external
surfaces under a well-defined contact angle. Rather the midpoint
contour $m(x,y,z)=0$ gradually bends over tangentially towards the
external surfaces, see Fig.~\ref{fig6}a. Some crowding of
contours near the external walls is always seen, implying that the
corrections discussed in Eqs.~(\ref{eq50})-(\ref{eq52}) will make
a substantial contribution \cite{66}. Also in the symmetric
situation (Fig.~\ref{fig6}b), where the interface is flat and not
bent, and the contour $m(x,y,z)=0$ does coincide with the basal
plane, one can see that the effective width of the interface is
quite broad. Due to this interfacial broadening, we expect that
the details of the singular shape of the system (external surfaces
with competing surface fields meet at $z=0$ under a sharp angle,
$2\alpha = 90^0$ here) do not matter, and if the bi-pyramid edges
would be rounded away by a smoothly curved behavior, we should 
still observe the same behavior as in the present study as long
as the radius of curvature in these smoothly curved regions is a
finite constant, independent of $L$.

We next turn to a Monte Carlo test of the free energy
constructions discussed in Fig.~\ref{fig3}. For this
purpose the surface free energy difference $f_s(H_s)$ is needed,
and in order to find this quantity we apply thermodynamic
integration methods \cite{10,LV}, as done in our recent study of wedge filling
\cite{30}.

Writing $f_{s+}(H_s),f_{s-}(H_s)$ for the surface excess free
energies of the bulk phases with positive (+) and negative (-)
magnetization and using the symmetry relation for the Ising model
$f_{s-}(H_s)=f_{s+}(-H_s)$, we find that the required surface free
energy difference can be written as
$f_s(H_s)=f_{s+}(H_s)-f_{s+}(-H_s)$. Recalling the relation from
surface thermodynamics of ferromagnets \cite{67}
\begin{equation}\label{eq54}
M_{s+}(H_s)=-(\frac {\partial f_s}{\partial H_s})\;,
\end{equation}
where $M_{s+}(H_s)$ is the local magnetization per spin in the
surface plane of an Ising ferromagnet with positive magnetization
in the bulk, subject to surface field $H_s$, we
recognize that the required free energy difference can be written
as
\begin{equation}\label{eq55}
f_s(H_s)= \int _{-H_s}^{H_s}M_{s+}(H'_s)dH'_s
\end{equation}
\begin{figure}
\includegraphics{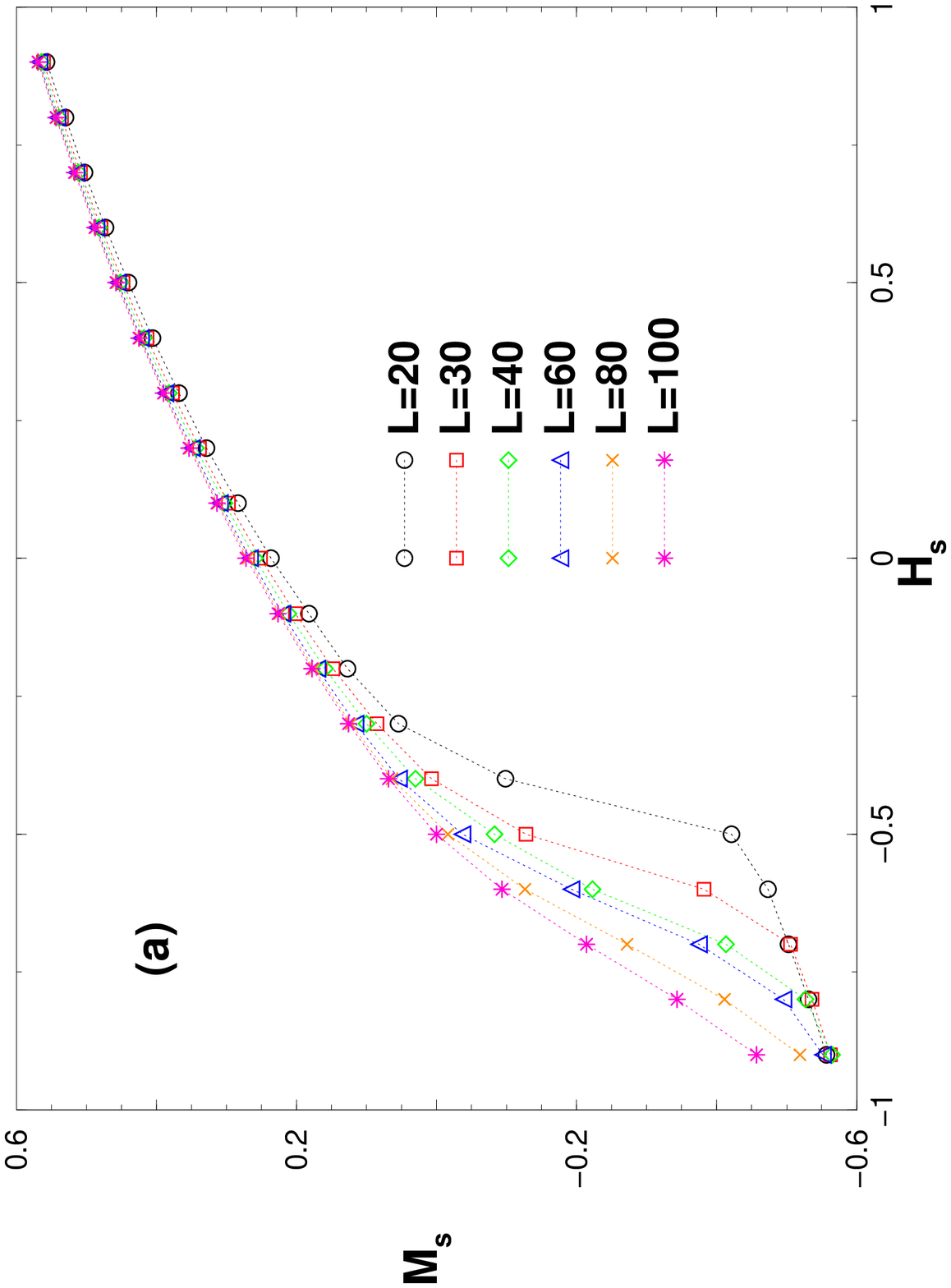}
\includegraphics{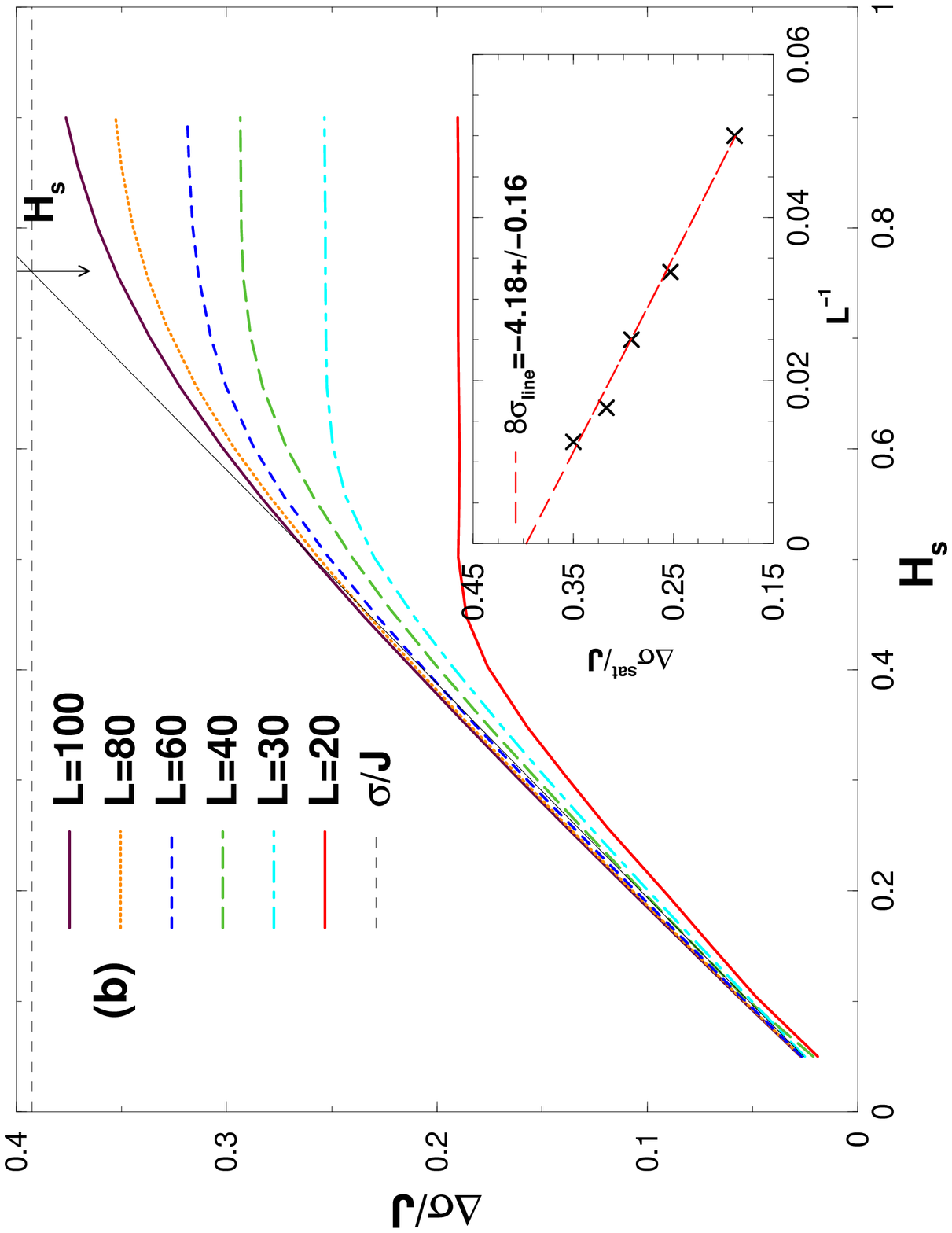}
\includegraphics{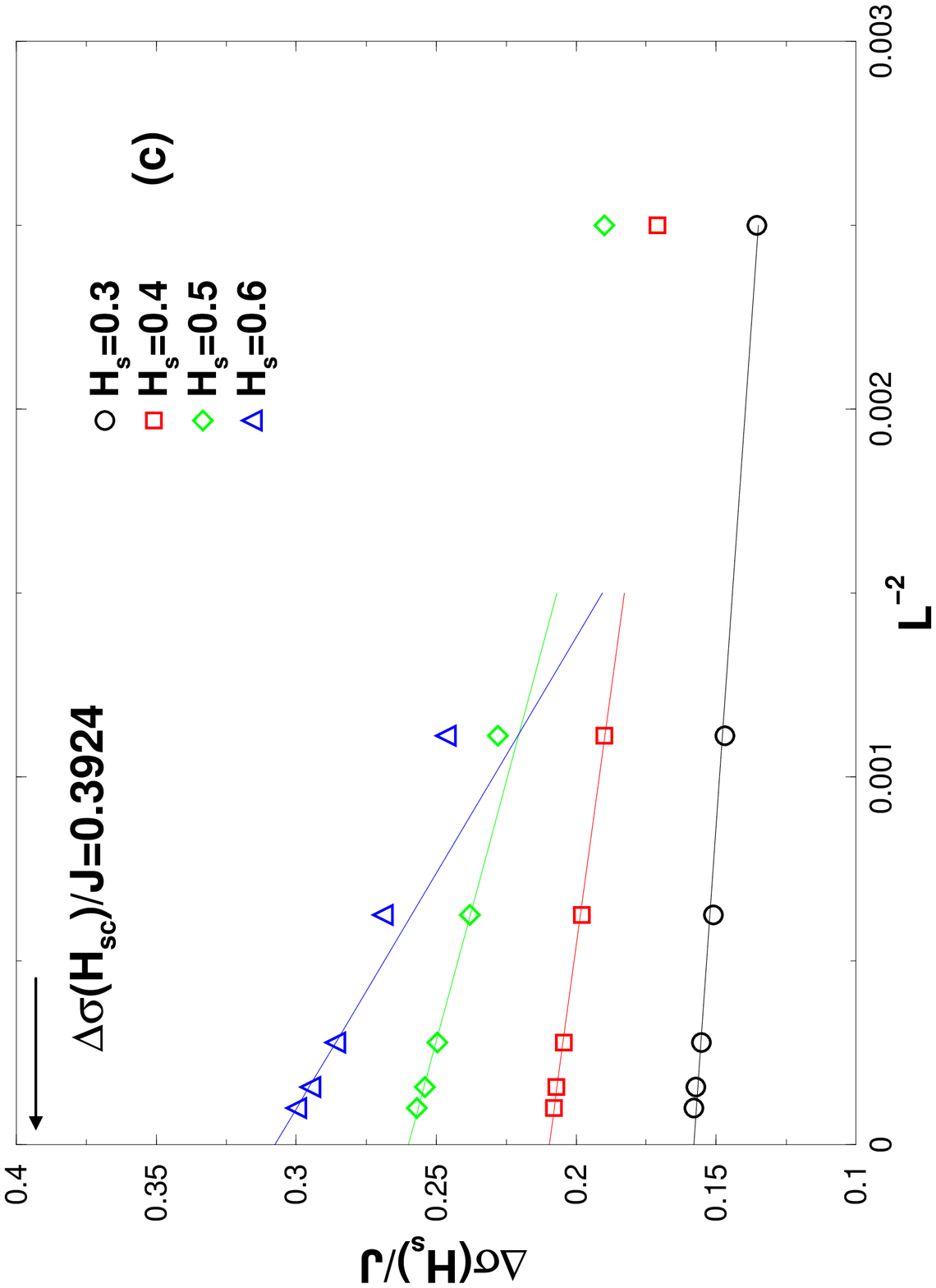}
\vskip 13.0cm
\caption{\label{fig7}(a) Local magnetization per spin,
$M_s$, in the surface plane [traditionally this quantity
\cite{67} is denoted as $M_1$, to avoid confusion with the surface
excess contribution $m_s$, to the total magnetization, cf.~Eq.~(\ref{eq50})], 
plotted versus the local surface magnetic field,
using parameters $k_BT/J=4$, $J_s/J=1/2$, and linear dimensions
$L$ in the range $20 \leq L \leq 100$, as indicated.
(b) Surface free energy difference $\Delta \sigma(H_s)/J$ plotted
vs surface magnetic field $H_s$, as obtained from the data in part
(a) via thermodynamic integration, Eq.~(\ref{eq55}). The full
straight line shows the result of an extrapolation of $F_s(H)/J$
to the thermodynamic limit (see part c). Broken horizontal
straight line marks the value\cite{68} of the interfacial tension, $\sigma$,
of the Ising model at $k_BT/J=4$. The inset
shows the extrapolation of the apparent plateau values (reached at
$H_s=0.9)$ of $\Delta \sigma /J$ versus $L^{-1}$. Arrow on top
shows $H_{sc}=0.76$. 
(c) Finite size extrapolation of the surface free energy
difference $\Delta \sigma (H_s)/J$ plotted vs. $L^{-2}$, for four
different choices of $H_s$, as indicated. Arrow shows the expected
value of $\Delta \sigma (H_{sc})/J$ at the phase transition.}
\end{figure}

Fig.~\ref{fig7}a shows a plot of $M_s$ versus $H_s$ for the Ising
bi-pyramids with various linear dimensions $L$ as used in the
present study. In principle, for an accurate estimation of
$f_s(H_s)$, one should use not bi-pyramid surfaces but rather
surfaces of large parallelepipeds, (oriented with the same angle
$\alpha$ relative to the plane $z=0$ as studied here), where
effects due to edges and corners could be avoided by using
suitably ``staggered'' periodic boundary conditions. Such a study
has not been attempted here, rather we work with the bi-pyramid
geometry throughout, but we then have to carefully consider the
finite size effects, which are indeed quite pronounced
(Fig.~\ref{fig7}b,c). Here $\Delta \sigma(H_s)=\sqrt{2}f_s(H_s)$,
to account for the fact that the total surface area of the
triangular facets of the pyramid is $\sqrt{2}(2L+1)^2$ in our model, while
the total surface area of the basal plane is $(2L+1)^2$, measuring
lengths in units of the lattice spacing. The corresponding value
of the interface tension of a planar interface in the Ising model
at $k_BT/J=4$, $\sigma/J=0.3924$ \cite{68}, is indicated by a
dashed horizontal line. In principle, we expect that $\Delta
\sigma /J$ near $H_{sc}$ is a straight line, which intersects
$\sigma/J$ precisely at $H_{sc}$. In fact, the values of $\Delta
\sigma (H_s)$ extrapolated to the thermodynamic limit do show such
a behavior, yielding $H_{sc}=0.76$. However, for all finite $L$
the curves $\Delta \sigma /J$ smoothly bend over, and reach
horizontal plateaus for large $H_s$, which are substantially lower
than $\sigma/J$. In terms of Fig.~\ref{fig7}b, the occurrence of
these plateaus is understood from the fact that $M_s$ for
sufficiently negative $H_s$ obeys already the symmetry
$M_s(H_s)=-M_s(-H_s)$, because the sign of the magnetization in
the corresponding pyramid has changed when the interface has moved
towards the basal plane. For fields $H_s$ where this symmetry
holds the integral in Eq.~(\ref{eq55}) yields vanishing further
contributions, resulting in a horizontal variation of $F_s(H_s)$ with
$H_s$ in this regime. Qualitatively, the behavior seen in
Fig.~\ref{fig7}b, closely resembles the expected behavior as
hypothetically sketched in Fig.~\ref{fig3}. The linear
extrapolation of the saturation plateaus $\Delta \sigma^{sat}/J$
versus $L^{-1}$ (inset of Fig.~\ref{fig7}b) is nicely consistent
with this picture, since the linearity of the plot asserts that
the depression of the plateaus indeed is a line tension effect,
and the extrapolated value ($\Delta \sigma ^{sat}/J\approx 0.39)$
agrees with $\sigma/J$ within the statistical error. From the
slope of the broken straight line in the inset in
Fig.~\ref{fig7}b we deduce the estimate
\begin{equation}\label{eq56}
8 \sigma ' _{\rm line}/J=-4.18 \pm 0.16\;.
\end{equation}
Unfortunately, we are not aware of any estimates of $\sigma '
_{\rm line}$ for our geometry in the literature, to which our result
could be compared.

\begin{figure}[!ht]
\includegraphics{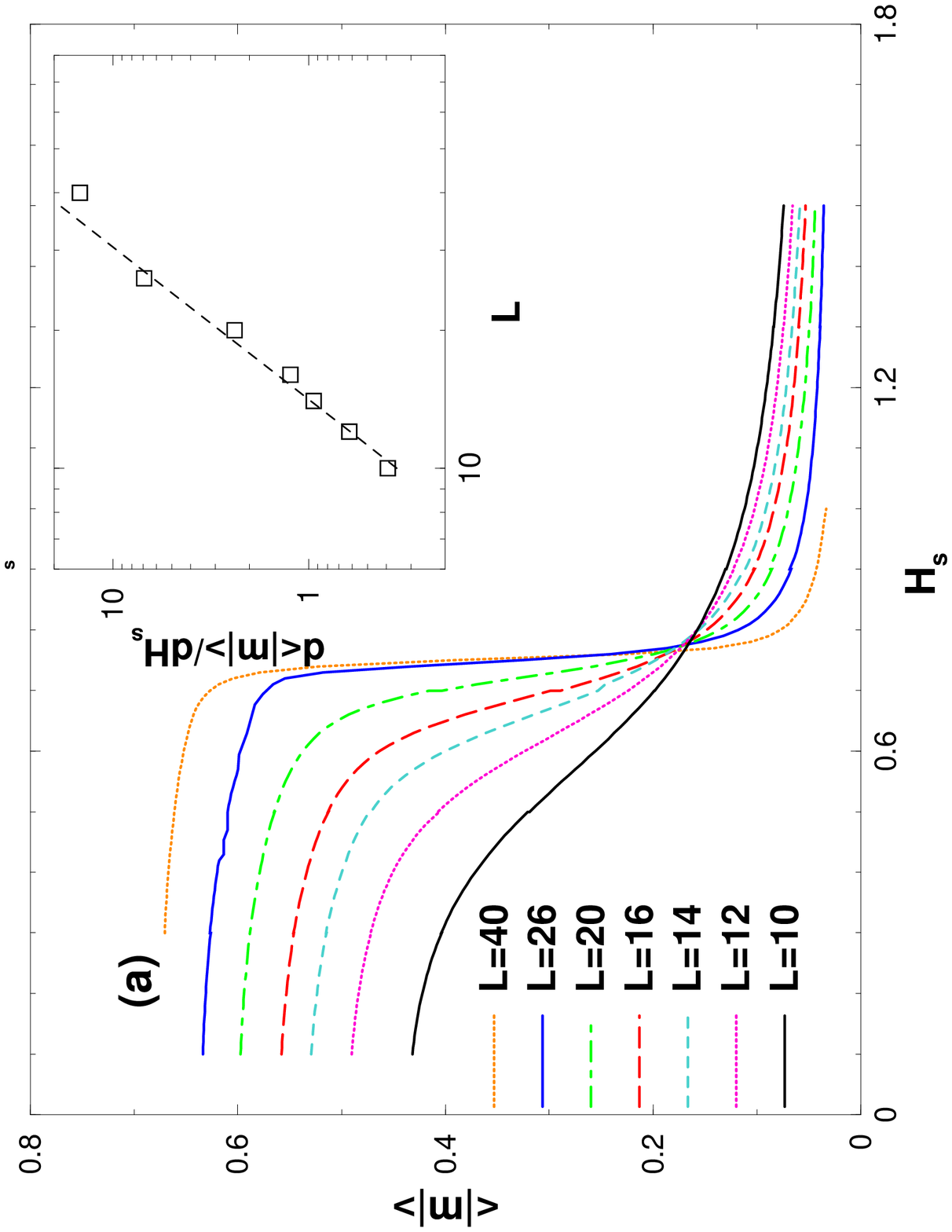}
\includegraphics{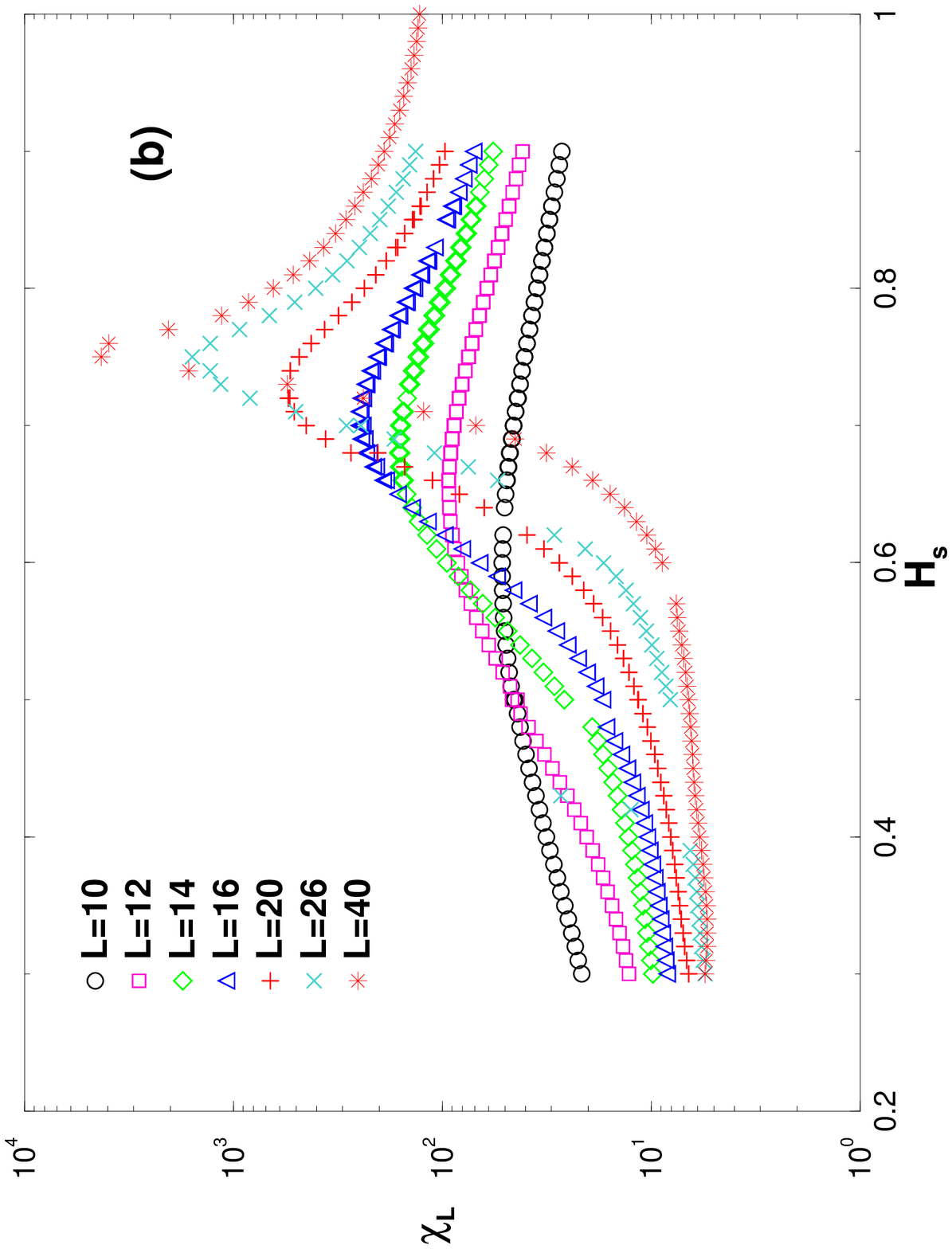}
\includegraphics{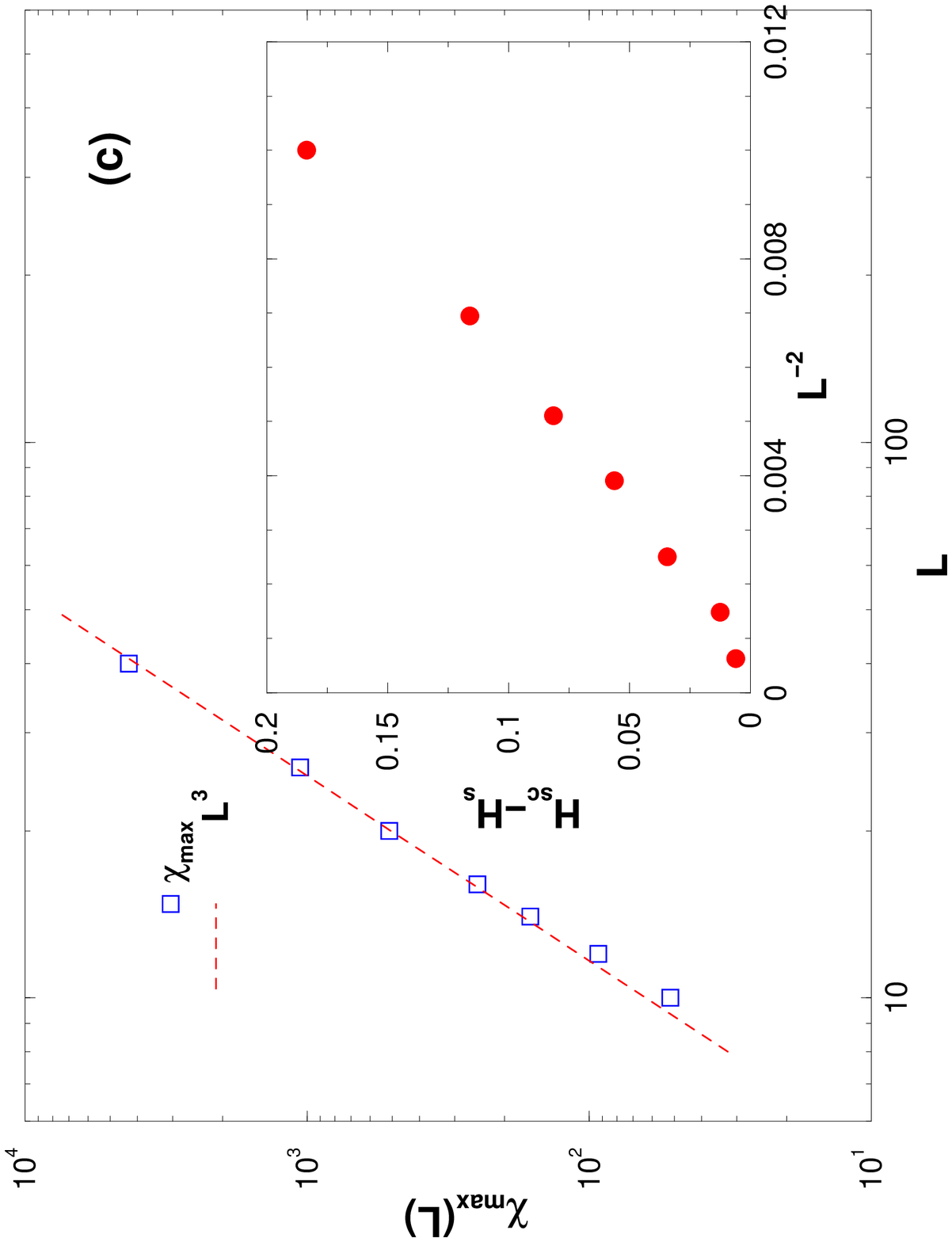}
\includegraphics{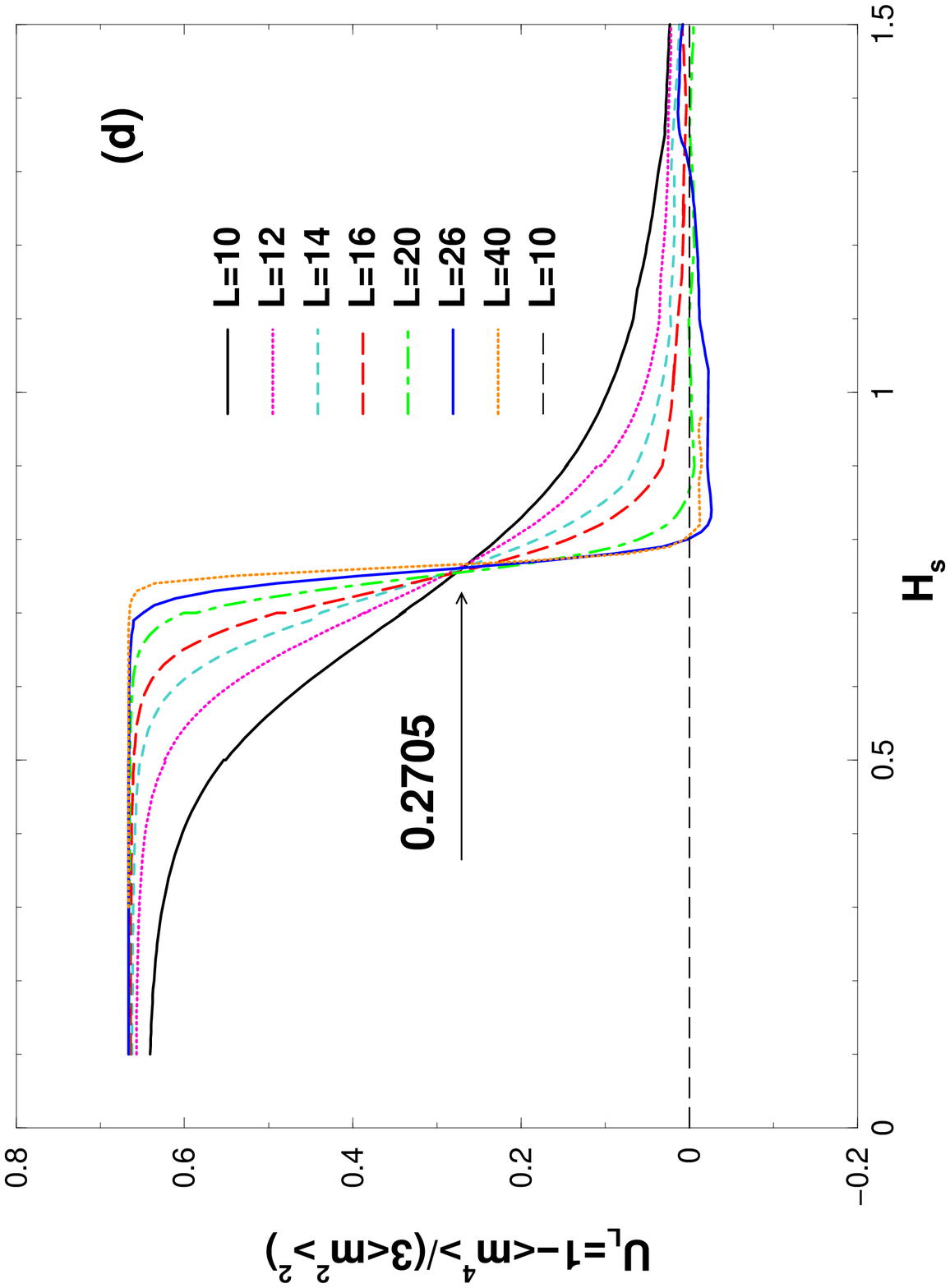}
\vskip 13.9cm
\caption{\label{fig8} (a) Plot of the absolute value
$\langle |m| \rangle$ of the magnetization of the Ising bi-pyramid
versus the surface magnetic field $H_s$, for $k_BT/J=4.0$, $H=0$,
$J_s/J=0.5$, and various linear dimensions L in the range $ 10
\leq L \leq 40$, as indicated. Inset shows a log-log plot of the
slope at the common intersection point vs. $L$. The straight line
illustrates the theoretical value of the slope. (b) Plot of the
susceptibility, calculated from magnetization fluctuations as
$k_BT\chi_L=L^3(\langle m^2\rangle - \langle |m|\rangle ^2)$, as a
function of the surface field $H_s$, including various linear
dimensions $L$ in the range $10 \leq L \leq 40$, as indicated.
System parameters are the same as in part (a). Note the
logarithmic scale of the ordinate. (c) Log-log plot of the
susceptibility maximum, $\chi_{max}(L)$, versus linear dimension,
for the systems shown in part (b). Broken straight line
illustrates the expected relation $\chi _{max}(L) \propto L^3$.
Inset shows the location of the maximum of $\chi _L,\;H_s^{max}$,
plotted vs $L^{-2}$, to illustrate the convergence of $H_s^{max}$
towards $H_{sc}$ as $L \rightarrow \infty$. (d) Cumulants $U_L$
(Eq.~(\ref{eq57})), plotted vs $H_s$, for various $L$ as shown in
the figure, for the same system parameters as used in parts (a) and
(b).}
\end{figure}
In the regime where $H_s$ is small, so that the bi-pyramid has
essentially a uniform magnetization, apart from the region close
to the lower pyramid corner (cf.~Fig.~\ref{fig6}a), the finite size
correction to the surface free energy difference $\Delta
\sigma(H_s)$ varies proportional to $L^{-2}$, as expected for a
corner correction (Fig.~\ref{fig7}c). Unfortunately, a reliable
extrapolation of $\Delta \sigma(H_s)$ in the region near $H_{sc}$
would require to simulate much larger systems than was possible to
us.

We now turn to the description of the phase transition in terms of
the moments $\langle |m|^k\rangle$ of the distribution function
$P_L(m)$ of the magnetization (Fig.~\ref{fig8}). Note that here
and in the following $m$ is the magnetization per spin and not
normalized by $m_b$. A striking fact is the common intersection
point of the $\langle |m|\rangle$ versus $H_s$ curves (part a) for
a broad range of choices for $L$, at $H_{sc}=0.76 \pm 0.005$. This
value of the intersection point is fully in accord with the
estimate resulting from the free energy intersections, obtained in
Fig.~\ref{fig7}b. The inset illustrates the fact that the slope
of the curves $\langle |m|\rangle$ vs. $H_s$ at $H_s=H_{sc}$
increases dramatically with $L$; in fact, the data are roughly
compatible with the behavior $(d\langle |m|\rangle
/dH_s)_{H_{sc}}\propto L^2$, that one immediately derives from the
scaling description, Eq.~(\ref{eq44}). Such a rapid increase of
the slope $(d\langle |m|\rangle/dH_s)_{H_{sc}}$ is rather uncommon
for normal second order transitions. From the curve for $L=40$
it is already easy to guess the limiting behavior, namely $\langle
|m|\rangle=m_b\approx 0.75$ for $H_s <H_{sc}$, while $\langle
|m|\rangle=0$ for $H_s >H_{sc}$. Nevertheless, this jump of $\langle |m|\rangle
$ resulting in the thermodynamic limit should not be mistaken for
a standard first order transition -- rather one deals here
with the limiting case of a second order transition, where the
critical amplitude diverges, and hence the critical region is
exceedingly narrow.
Fig.~\ref{fig8} demonstrates that the susceptibility develops a
sharp peak of rapidly increasing height, as $L$ increases. As
expected from the Curie-Weiss law with the divergent critical
amplitude, the curves do not settle at a common $L$-independent
function away from $H_s=H_{sc}$. But the data clearly indicate a
gradual growth of $\chi_L$ with $H_s$ as $H_{sc}$ is approached
from either side of the transition, and the width over which this
peak is rounded rapidly shrinks as $L$ is increased. There is not a
convergence to a delta function singularity, that would
characterize a standard first order transition \cite{40,41}.

A direct analysis of these data, not requiring any bias from
theory, examines the growth of the peak height with $L$, and the
scaling of the peak position $H^{max}_s - H_{sc}$ with $L$
(Fig.~\ref{fig8}c). One nicely recognizes that also these data are
compatible with a transition at $H_{sc}\approx 0.76$, and the
relation $\chi_{max}(L) \propto L^3$ implies that the maximum
values of $\langle m ^2\rangle-\langle |m|\rangle ^2 $ are of
order unity, as expected on the basis of Eq.~(\ref{eq44}). Unlike
first order transitions (where also the susceptibility peak height
increases proportional to the volume) the width of the
susceptibility peak does not shrink to zero for $L\rightarrow
\infty$.

A standard method to locate critical points for various phase
transitions is to check for intersections of the reduced fourth
order cumulant \cite{41,42,43},
\begin{equation}\label{eq57}
U_L=1-\langle m^4 \rangle/(3 \langle m^2\rangle ^2)\;.
\end{equation}
Plotting hence $U_L$ vs. $H_s$ for various $L$ should yield a
common intersection point at $H_s=H_{sc}$. Fig.~\ref{fig8}d shows
that this simple recipe works here rather well again, confirming
the previous estimate $H_{sc}\approx 0.76$. If one accepts Eq. (\ref{eq45})
as a description of the distribution at $H_{sc}$, one predicts for the
value of $U^*$ of the cumulant at the intersection point the value
$U^*=1-\Gamma(5/4)\Gamma(1/4)/[3\Gamma(3/4)^2]\approx 0.2705$. The
arrow in Fig.~\ref{fig8}d shows that this value is in very good 
agreement with the data.  Interestingly, these
\begin{figure}
\includegraphics{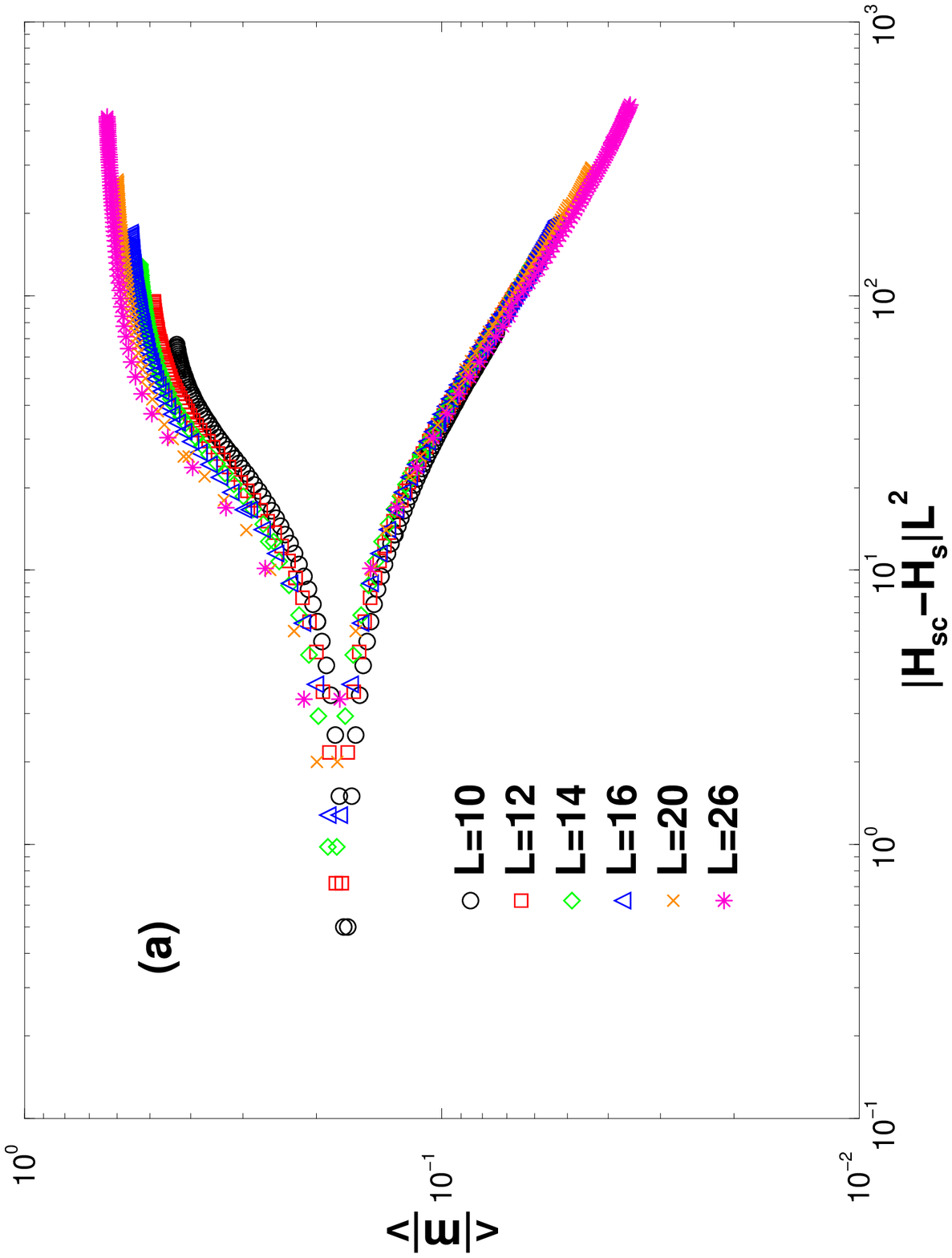}
\includegraphics{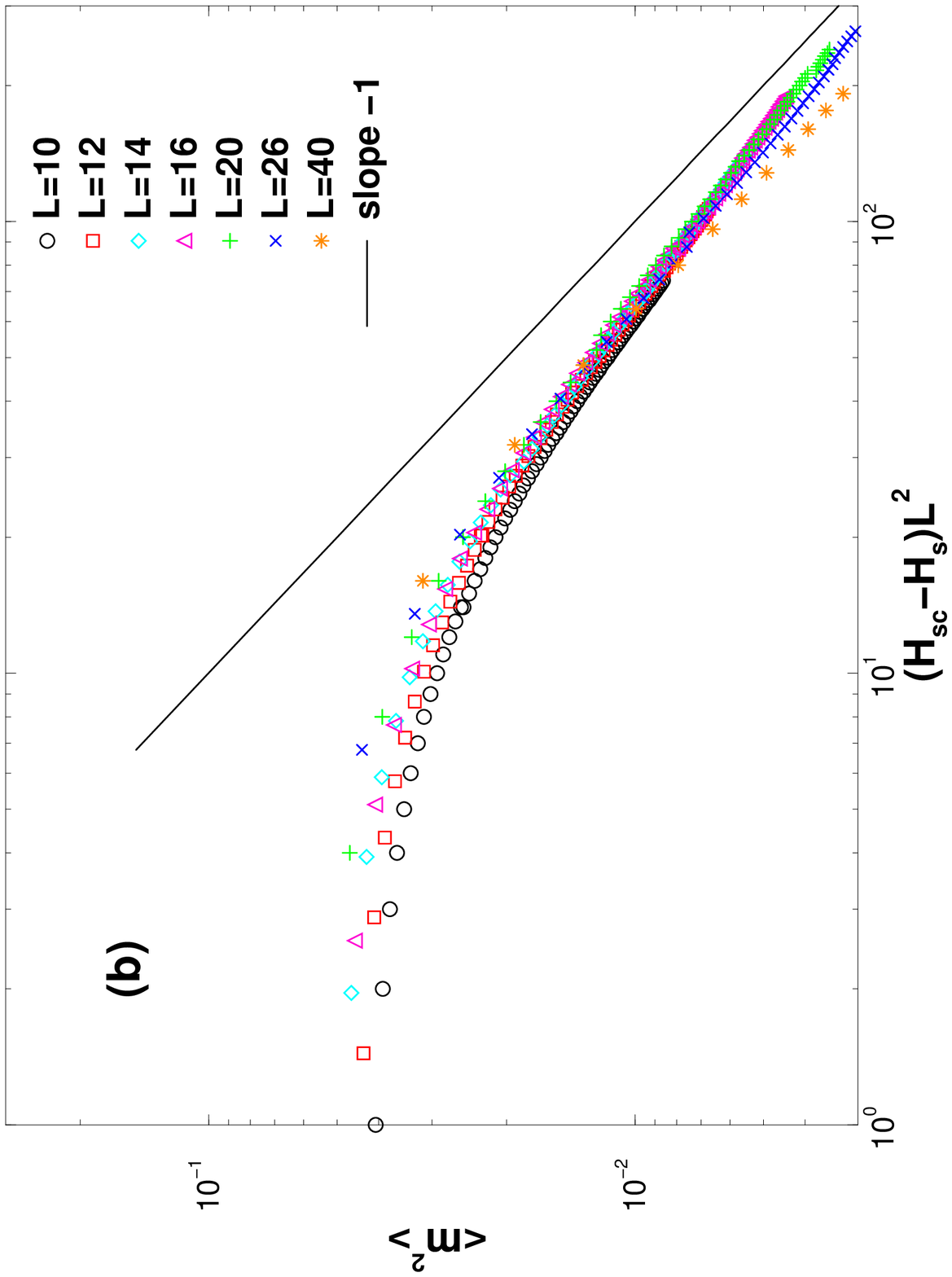}
%\special{psfile=fig9c.ps hoffset= 90 voffset=-180 hscale=35 vscale=35 angle=270}
\vskip 7.5cm
\caption{\label{fig9} (a) Absolute value of the
magnetization $\langle |m|\rangle$ plotted vs. the scaling
variable $|H_s-H_{sc}|L^2$, using the data of Fig.~\ref{fig8}, and
$H_{sc}=0.76$ and various L as indicated in the figure. \\ (b)
Plot of the second moment, $\langle m ^2 \rangle$, vs. the scaling
variable $(H_s-H_{sc})L^2$, using $H_{sc}=0.76$ and including data
only for $H_s>H_{sc}$. Various choices of L are shown as
indicated. The straight line shows the slope of the asymptotic
Curie Weiss law, $\langle m ^2 \rangle \propto (tL^2)^{-1}$. 
}
\end{figure}
data develop not only a common intersection point, but also a
shallow minimum for $H_s >H_{sc}$. This is somewhat reminiscent of
the behavior at thermally driven first order transitions, such as
occur in the $q$-state Potts model in $d=3$ dimensions for
$q \geq 3$, where $U_L$ is known to exhibit a very deep minimum
{$U_L^{min}\propto - L^3 \quad$ \cite{69}}. So the behavior of the
cumulant is again indicative of a second order transition that is
close to a first order transition.

\begin{figure}[h]
\includegraphics{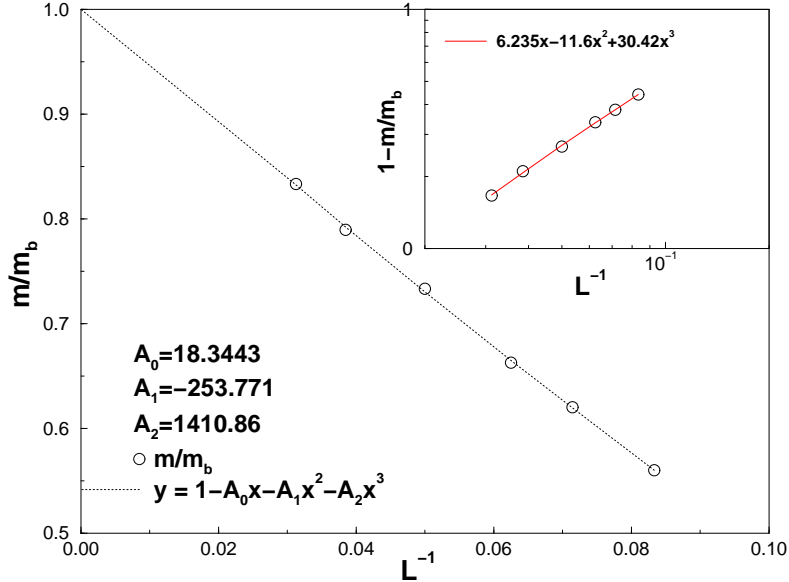}
\vskip 7.5cm
\caption{\label{fig10} Position of the maximum of
the distribution $P_L(m)$ at $H_s=0.73$ plotted vs. $L^{-1}$, for
the parameters $k_BT/J=4.0$ and $J_s/J=1/2$. The bulk
magnetization is $m_b=0.750$. The numerical data are fitted to
Eq.~\ref{eq51}, and the constants $A_0,A_1,A_2$ are quoted in the
figure. Inset shows the difference $1-m/m_b$ vs. $L^{-1}$ on a
log-log plot.}
\end{figure}

We now turn to a more detailed test of the finite size scaling
predictions, in particular of Eq.~(\ref{eq44}). Fig.~\ref{fig9}
shows scaling plots of the magnetization and the magnetization
square. Using $|H_{sc}-H_s|L^2$ as scaling
variables in part (a), both branches of the scaling
functions for $H_s <H_{sc}$ (upper branch) and $H_s>H_{sc}$ (lower
branch) are combined in a single plot. However, the ``data
collapsing'' on master curves in parts (a) and (c) is not really
perfect, and some corrections to scaling are clearly seen.
However, as pointed out above, \{see Eqs.~(\ref{eq50})-(\ref{eq52})
and the accompanying discussion\}, our scaling description has ignored
contributions such as due to the surface excess magnetization
$m_s$. Fig. ~\ref{fig10} shows a fit of data for the magnetization
to Eq.~(\ref{eq51}) to test the size dependence, and thus it is
shown that indeed important corrections are present \cite{66}.

\begin{figure}[h]
\includegraphics{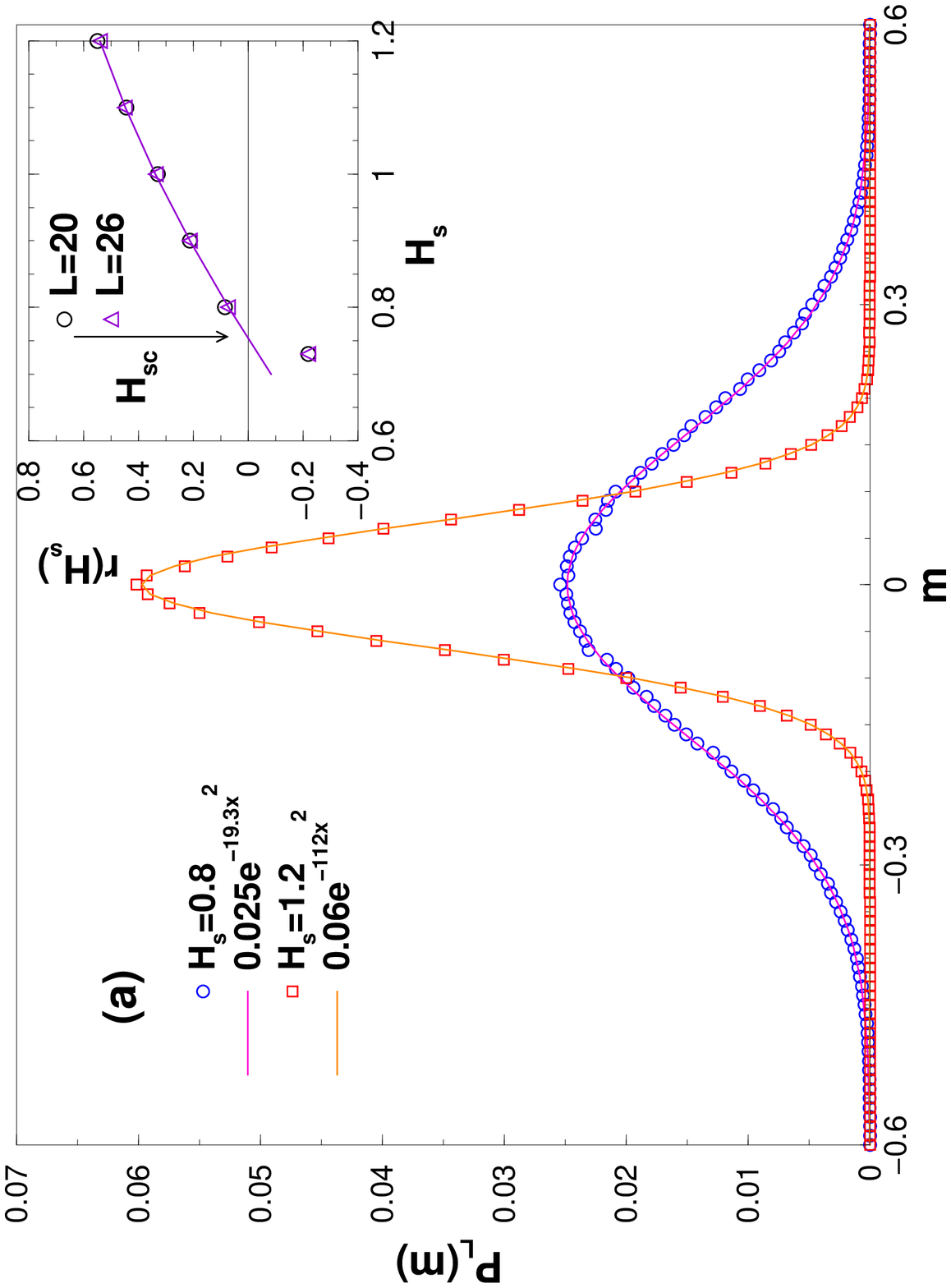}
\includegraphics{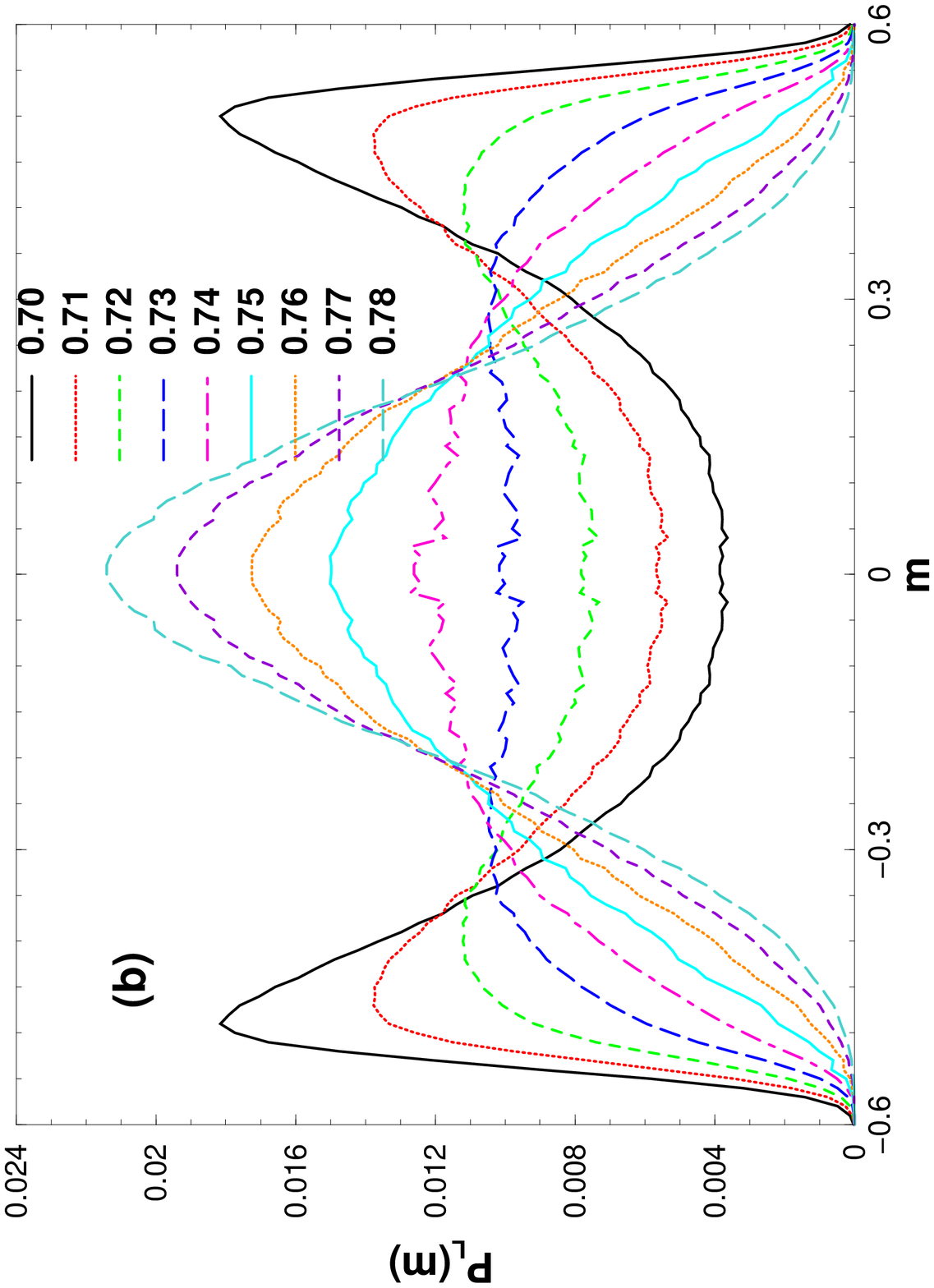}
\vskip 6.0cm
\caption{\label{fig11} (a) Probability distribution
of the magnetization $P_L(m)$ for fields $H_s >H_{sc}$, outside
the critical region, using the parameters $L=20$, $k_BT/J=4$,
$J_s/J=1/2$, and $H_s=0.8$ or $1.2$, respectively. Curves through
the data points show fits to simple Gaussian, as indicated in the
figure. The inset shows the function $r(H_s)$, obtained from such
fits (for $L=26$) to demonstrate the change of sign near $H_{sc}$.
The data points for $H_s=0.73$ are taken from Fig.~\ref{fig12}b
below. (b) Probability distribution $P_L(m)$ of the
magnetization, $m$, of an Ising bi-pyramid for $L=20$, $k_BT/J=4$,
$J_s/J=0.5$, and $H=0$. Curves show various surface fields
$H_s/J$, as indicated in the key.}
\end{figure}

The simulation data in Fig.~\ref{fig8} were extracted from an
analysis of the probability distribution $P_L(m)$ of the
magnetization, $m$, in the finite bi-pyramid, and since $P_L(m)$
plays a key role in our phenomenological description (Sec. II), we
discuss $P_L(m)$ in detail now. Fig.~\ref{fig11}a shows that in
the phase where the interface coincides with the basal plane of
the bi-pyramid, $P_L(m)$ is perfectly described by the simple
Gaussian, Eq.~(\ref{eq27}). From Fig.~\ref{fig8}d we have already
seen that the fourth order cumulant rapidly tends to zero for
$H>H_{sc}$ as $L \rightarrow \infty$. The very good Gaussian fits
of Fig.~\ref{fig8}a imply that all higher order cumulants vanish
as well. Of course, this behavior is plausible due to the rapid
decrease of the fourth order term $u_Lm^4/4$ in Eq.~(\ref{eq31})
as $L \rightarrow \infty$, since $u_L\propto L^{-3}$. Thus, for
$(H_s-H_{sc})L^2 \gg 1$ the second moment $\langle m^2 \rangle$
shown in Fig.~\ref{fig9}b already contains the full information on
the distribution.

\begin{figure}[h]
\includegraphics{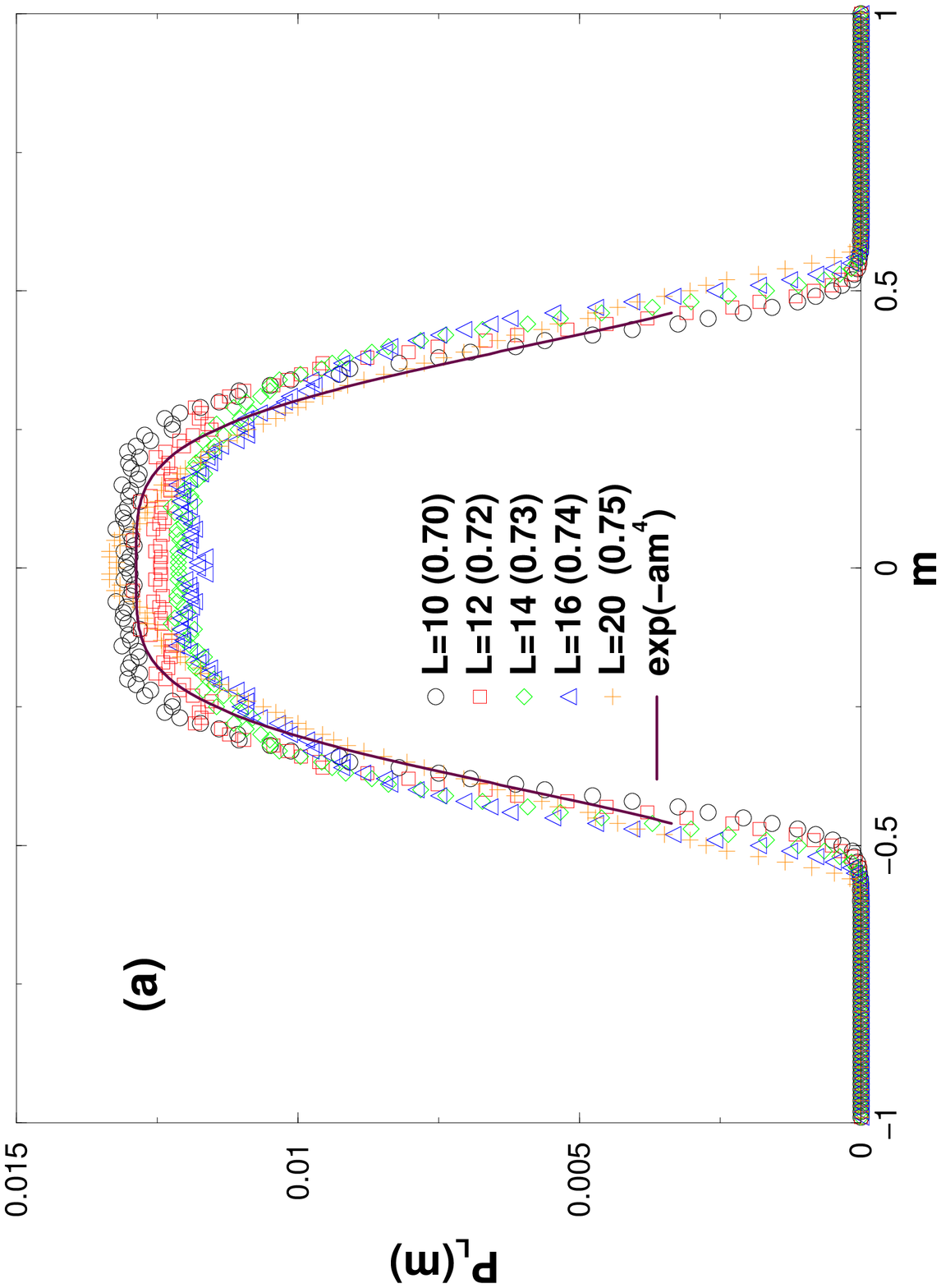}
%\special{psfile=fig12b.ps hoffset=220 voffset=35 hscale=39 vscale=39 angle=270}
\includegraphics{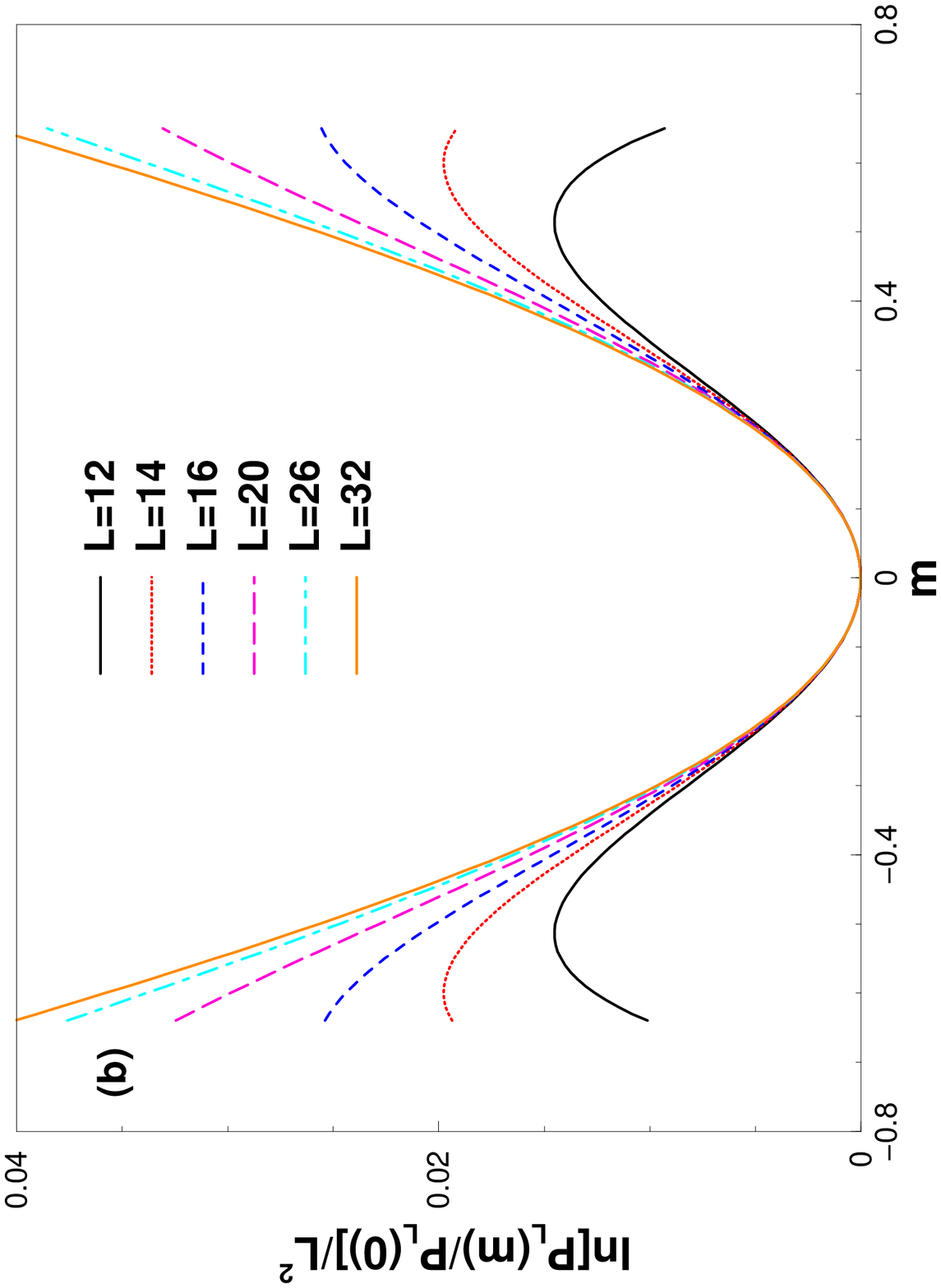}
\vskip 7.5cm
\caption{\label{fig12} (a) Probability distribution
$P_L(m)$, of the magnetization, $m$, of Ising bi-pyramids for
$k_BT/J=4,\; J_s/J=0.5,\; H=0$, and various choices of $L$ with
accompanying choices of $H_s/J$ for which a flat variation of
$P_L(m)$ near $m=0$ was expected (these choices are quoted in the
figure). Full curve shows the theoretical variation from
Eq.~\ref{eq45}, $P_L(m) \approx \exp [-am^4]$, with
$a=2u/(3k_BT)=(1/3)u/J\approx 30.4$. \\
(b) Plot
of $\ln [P_L(m)/P_L(0)]/L^2$ vs. $m$, for $k_BT/J=4,\; J_s/J=0.5$, $H=0$, at fixed $H_s/J=0.73$ and
various $L$. 
The quadratic part near $m=0$ is described by $\ln
[P_L(m)/P_L(0)]/L^2=0.11 m^2$ independent of $L$.}
\end{figure}

Fig.~\ref{fig11}b demonstrates now the smooth change of $P_L(m)$
from the single Gaussian to the double Gaussian when for fixed L
the strength of the surface field is varied. This behavior is
fully in accord with expectations for second order phase transitions. For a
first order transition (e.g., a first order interface
localization-delocalization transition has been recently studied
both for an Ising system in thin film geometry with
$J_s/J=1.5,H_s/J=0.25$ \cite{70} and for models of confined
polymer mixtures \cite{57,72}) the corresponding distribution
$P_L(m)$ near the transition has a pronounced three-peak
structure: two peaks have nonzero positive or negative
magnetization, and the third peak occurs for $m=0$ (e.g. see
Fig.~\ref{fig8}c of Ref.~\cite{70} for an explicit example.). In
contrast, here we expect near the transition a single very flat
and broad peak (whose width should not shrink with increasing
linear dimension $L$, as emphasized in Eq.~(\ref{eq45})).
Fig.~\ref{fig12} therefore examines the size dependence of
$P_L(m)$ in the critical region, and part (a) shows that indeed one
can find for each $L$ a field $H_{sc}'(L)$ such that $P_L(m)$ is
essentially flat near $m=0$, and approximately the width of
$P_L(m)$ stays independent of $L$ when $L$ increases, while
$H_{sc}'(L)\rightarrow H_{sc}=0.76$ as $L$ increases.

Ideally, one might have expected that Eq.~(\ref{eq45}) should hold
strictly for $H_s=H_{sc}$ (i.e., $t=0$). However, the small
variation of $H'_{sc}(L)$ with L that is implied by
Fig.~\ref{fig12}a does not invalidate our phenomenological theory
of Sec. II at all: as is well known \cite{38,39,40,41,42,43} for
finite systems there is no unique ``pseudocritical'' point, due to
the finite size rounding different criteria to locate a
``pseudocritical'' point for a finite system yield results
differing from each other (and from the true location of the
critical point) by amounts which are of the same order as the
rounding, i.e. $t \propto L^{-2}$ in our case. Such an argument
would imply $H'_{sc}(L)-H_{sc}\propto L^{-2}$ here. Unfortunately,
our simulation data are not accurate enough to check this relation
(and also a larger range of linear dimensions $L$ than what is 
available for Fig.~\ref{fig12}a would be desirable).

If a more rapid variation of $H'_{sc}(L)-H_{sc}$ than the second
power of $1/L$ results, it could be attributed to corrections to
finite size scaling, some of which were identified above
[Eqs.~(\ref{eq50})-(\ref{eq52})]. Despite all shortcomings that
our numerical results still have, we consider Fig.~\ref{fig12}a as
a highlight of the present study, since it demonstrates that in
the limit $L \rightarrow \infty$ at the transition point
\textit{macroscopic fluctuations} occur, the magnetization varies
essentially everywhere in the region from $m=-0.5$ to $m= + 0.5$
(in a situation where the bulk magnetization is $m_b \approx
0.75$), because the interface can correspondingly move freely up
and down. Of course, viewing the Monte Carlo simulation as a
stochastic process, such interface ``motions'' are extremely slow,
and this ``critical slowing down'' \cite{73} hampers severely the
statistical accuracy of our Monte Carlo study, as expected
\cite{74,75}. Note that we have applied single spin flip Monte
Carlo algorithms here, since the Swendsen-Wang algorithm
\cite{74,75} or related cluster algorithms are not offering any
advantage in our case, working for temperatures distinctly below
the bulk critical temperature $T_{cb}$ and in the presence of
nonzero surface fields.

Fig.~\ref{fig12}b shows then the magnetization distribution
$P_L(m)$ for various $L$ at a fixed value of $H_s$ that is
definitely below $H_{sc}$. One can see that with increasing values
of $L$ pronounced peaks develop with a very deep minimum in
between; in fact, it was necessary to apply so-called
``multicanonical'' sampling technique (see. e.g.~Refs.~\cite{75,76}) in
order to be able to sample more than $15$ orders of magnitude in
probability with sufficient accuracy.

We also remark that Fig.~\ref{fig12}b is of a very different
character than the corresponding distribution for a bulk Ising
system for $T<T_{cb}$ \cite{76}: there also a deep minimum in
between the peaks corresponding to the two signs of the order
parameter occurs, but it is very flat, almost horizontal, due to
configurations described as two phase coexistence (e.g., in a bulk
Ising hypercube with periodic boundary conditions slab-like
domains occur). Here the minimum of $P_L(m)$ near $m=0$ does not
correspond to a ``mixed state'' of the degenerate phases (the
interface being bound either to the top or to the bottom corner of
the bi-pyramid, respectively), since no such ``mixed state'' can
exist. Rather, the minimum corresponds to a uniform displacement
of the interface from its stable position near one of the two
corners to the basal plane.

Therefore the logarithm of the distribution $\ln P_L(m)$ near
$m=0$ is a simple parabola, as expected from Eqs.~(\ref{eq30}),
(\ref{eq31}). This is demonstrated clearly 
because we have normalized $\ell n P_L(m)$ such
that all minima coincide. Dividing out the predicted $L^2$
dependence, we see that all curves near the minimum nicely
\begin{figure}[h]
\includegraphics{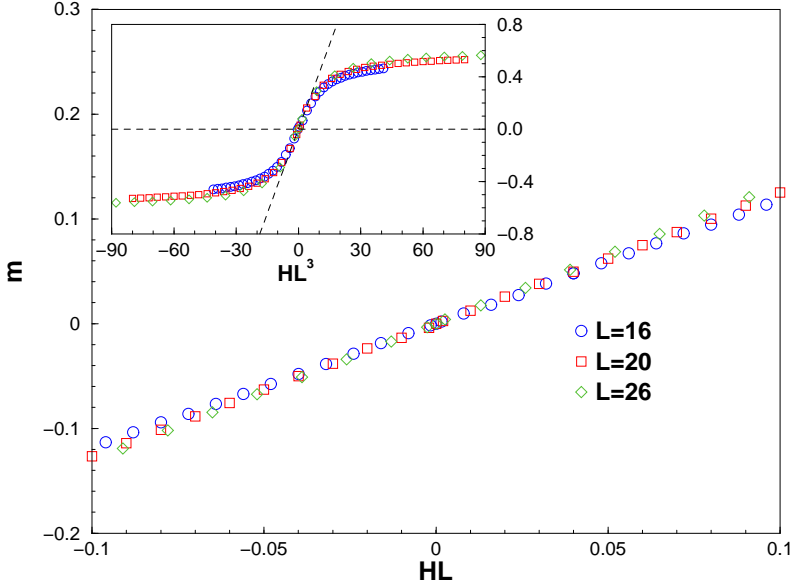}
\vskip 7.5cm
\caption{\label{fig13} Plot of the average
magnetization $\langle m \rangle$ as a function of the scaled
field $HL$, for the parameters $k_BT/J=4,\; J_s/J=0.5,\;
H_s=1.2$, and three choices of L as indicated. Inset shows
$\langle m \rangle $ vs. $HL^3$ at $H_{sc}=0.76$.}
\end{figure}
superimpose. However, it is also clear from Fig.~\ref{fig12}b that
a fit to the form
\begin{equation}\label{eq58}
\frac{\ln [P_L(m)/P_L(0)]}{8L^2/(3k_BT)}= -\frac 1 2 r m ^2 -
\frac{u}{4L^2} m^4\;,\quad H=0\;,
\end{equation}
which would be implied by Eqs.~(\ref{eq30}), (\ref{eq31}), when
$u_L=u/L^3$ is used, is not a good representation of the data for
large $m$, and higher order terms (of order $m^6, m^8, \ldots$)
would be required. Of course, this is not surprising at all, since
the saturation value of the magnetization at the considered
temperature is $m_b=0.75$, and the distribution spans the range
from about $m \approx - 0.6$ to about $+0.6$, i.e. close to the
saturation values of $m$. Of course, Eq.~(\ref{eq58}) is supposed to
be valid only for $|m|\ll m_b$. If we ignore this problem, the
``best fit'' values of the data in Fig.~\ref{fig12}b would be
$-4r/3k_BT=0.11$, $2u/3k_BT=30$. Nevertheless it is reassuring that the
estimates for $r$ resulting from the fits in Figs.\ref{fig11},
\ref{fig12} ($8r/(3k_BT) = 0.56,0,0965, -0.22$, for $H_s=1.2, 0.8, 0.73$,
respectively) yield a smooth curve $r(H_s)$, which has a zero
close to $H_{sc}=0.76$, though this curve is clearly not a simple
straight line over the wide range of values for $H_s$ that is
considered here.
\begin{figure}[h]
\includegraphics{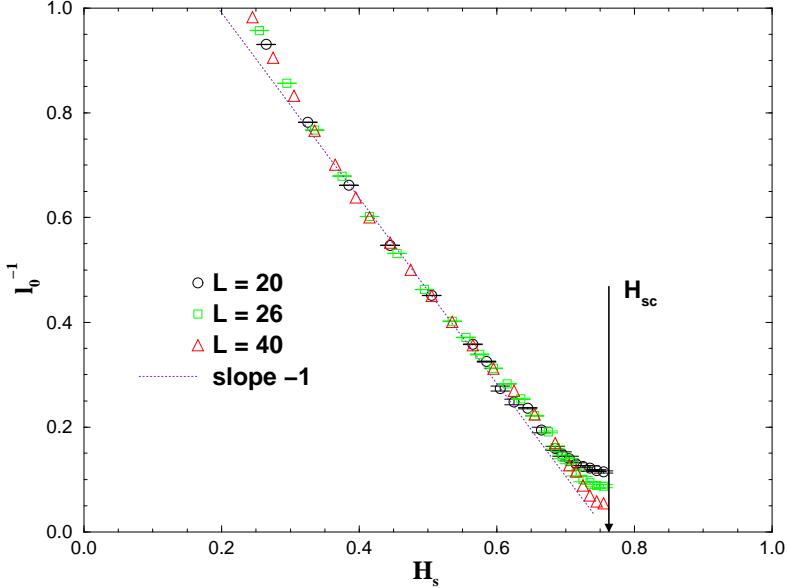}
\vskip 7.5cm
\caption{\label{fig14} Plot of the inverse distance,
$\ell _0^{-1}$, of the interface from the bottom corner vs. the
surface magnetic field $H_s$. Here $\ell_0 $ is obtained from an
analysis of the spatial magnetization distribution in the
bi-pyramid (cf.~Fig.~\ref{fig6}). Three lattice sizes are shown,
and the straight line marks the prediction $\ell _0 ^{-1} \propto
H_{sc}-H_s$ according to Parry et al. \cite{21}.}
\end{figure}

Now we turn, very briefly, to the behavior in non zero bulk
magnetic field $H$ (Fig.~\ref{fig13}). There are no surprises: we
find that $\langle m \rangle$ is a function of a scaled field $HL$
for $H_s > H_{sc}$, and of a scaled field $HL^3$ for $H_s=H_{sc}$,
as expected from the theory of Sec. II (cf. Eqs.~(\ref{eq26}),
 (\ref{eq46})). If we include in the Eq. (\ref{eq45}) the magnetic
field, $P_L(m)\propto \exp[-am^4+8L^3mH/3k_BT]$, we can readily calculate 
$m(H)$ at the critical point ($H_s=H_{sc}, t=0$) as $m(H) = a^{-1/4}
b[\Gamma(3/4)/\Gamma(1/4)]\{1-(b^2/2)[1-\Gamma(5/4)/(3\Gamma(3/4))]+ ...\}$,
where $b=8a^{-1/4}L^3H/(3k_BT)$. The resulting slope of $m(H)$ vs.
$L^3H/J$, $m(H)\approx 0.043L^3H/J$ is again in good agreement
with the numerical data in the inset of Fig.~\ref{fig13} and thus 
provides a test that the estimate of the constant $a$ ($a\approx
30 - 34$, see Fig.~\ref{fig12}a) is reasonable.

Finally, Fig.~\ref{fig14} considers the variation of the interface
distance when the interface is close to a corner, and hence the
theory of Parry et al. \cite{21} should straightforwardly apply
(i.e., Eq.~(\ref{eq4}) should hold). Thus we plot simply $\ell
^{-1}_0$ vs. $H_s$ to test the resulting linear variation.
Indeed one recognizes that the data are nicely compatible with the
predicted linear variation over a reasonable range of
$\ell^{-1}_0$, and the extrapolated intersection point with the
abscissa agrees well with $H_{sc}=0.76$. Of course,
we cannot expect that this relation works for $\ell_0^{-1} \geq
1$, since the notion of an interface becomes absolutely
meaningless when its distance from the pyramid corner becomes of
the order of a single lattice unit, or even less. Conversely,
finite size effects set in when $\ell_0 \approx L/2$. In view of
the fact, that the width of the critical distribution of the
interface fluctuating around the basal plane of the bi-pyramid is
of the order of $\Delta m \approx 0.5$, see Fig.~\ref{fig12}a, the
strong finite size effect seen in Fig.~\ref{fig14} is not at
all unexpected.

\section{DISCUSSION}

We start by summarizing the main findings of this study:

\begin{enumerate}
\renewcommand{\labelenumi}{\roman{enumi}.}
\item
The Ising ferromagnet in a geometry with free surfaces, where
surface fields are applied such that one half of the surface
experiences a positive surface field and the other half
experiences a negative surface field (of the same absolute
strength) has no total magnetization below the bulk critical
temperature $T_{cb}$ down to the filling transition temperature
$T_f(H_s)$, see Fig.~\ref{fig1}. This happens because the boundary
conditions stabilize the interface, separating two domains of
opposite orientation of the magnetization but taking equal volume.
Below $T_f(H_s)$, the interface (in the limit $L \rightarrow
\infty$) has essentially disappeared (it may be located at a
finite distance of order unity close to either the top or the
bottom corner if one specifically considers the bi-pyramid
geometry as done in Fig.~\ref{fig1}). 
Alternatively, this transition may be
driven by variation of the strengths of the surface field $H_s$
through its critical value $H_{sc}(T)$, note that $H_s=H_{sc}(T)$
simply is the inverse function of $T=T_f(H_s)$ in the $(T,H_s)$
plane. This transition occurs in the way
described here only if the parts of the surface where the
surface field has the same sign are all adjacent to each other.
E.g., for a bi-pyramid with boundary
condition where the sign of the surface field alternates
from one triangular surface to the adjacent one on the
same pyramid, this condition would be violated, and no such phase
transition occurs: rather one observes only a rounded phase
transition near the bulk transition point \cite{77}.

\item
As described in Fig.~\ref{fig3}, in the thermodynamic limit the
transition can be described in analogy with bulk first order
transitions, where an intersection of the two
branches of the free energy which describe the phases occurs.
However, here, both branches
are surface free energies, scaling like the surface
area ($\propto L^2$) rather than the volume ($\propto L^3$).
Although in the limit $L \rightarrow \infty$ the transition is
characterized by a discontinuous jump in the magnetization
(cf.~Figs.~\ref{fig1} and \ref{fig8}a), it nevertheless 
is a second order transition, if the line tension of the
boundaries of the interface where it meets the walls
(Fig.~\ref{fig2}, right part) is \textit{negative}. This negative
line tension makes it energetically favorable to stabilize
already a domain of the minority phase for
$H_s<H_{sc}$ with a mesoscopic linear dimension $\ell _0$, whereby
$\ell _0$ in the limit $L \rightarrow \infty$ diverges
continuously, $\ell _0 \propto (H_{sc}-H_s)^{-1}$. Thus
in a small interval of $H_s$ close to $H_{sc}$ in
Fig.~\ref{fig8}a, or in a small interval of temperature near
$T_f(H_s)$ in Fig.~\ref{fig1} (right part), the magnetization
$\langle |m| \rangle$ of the system varies continuously from its
saturation value $m_b$ to zero. In the thermodynamic limit,
however, the width of this interval shrinks to zero. On the other
hand,
this width over which the smooth variation occurs is of the same
order as the width of the interval over which the finite size
rounding of the transition occurs, namely of order $L^{-2}$.
Therefore the power law, Eq.~(\ref{eq33}), predicted by a simple
Landau-like theory that ignores finite size rounding, is
nowhere clearly observable. In the finite size scaling plot
(Fig.~\ref{fig9}a) one cannot identify a branch with a slope $1/2$ on
the log-log plot on which the curves collapse. This is
prevented by the saturation of the order parameter and, thus, the curves
bend over to flat plateaus. This saturation 
is not described by the theory we have developed here, and it
clearly violates the finite size scaling, as is evident from
Fig.~\ref{fig9}a.

\item
A particular interesting behavior exhibits the total
susceptibility of the system.
Figs.~\ref{fig8}b and \ref{fig9}b imply, in accord with our theory,
that the susceptibility $\chi$ shows a Curie-Weiss-like divergence
for $H_s>H_{sc}$, Eq.~(\ref{eq26}). The region of bulk fields,
where this divergence is observable, shrinks to zero like $1/L$,
because the critical amplitude in Eq.~(\ref{eq26})
varies like $L$.  There is a remarkable asymmetry between the behavior of the
susceptibility in the regime $H_s>H_{sc}$ (where no total
magnetization occurs) and the regime $H_s<H_{sc}$, however: in the
latter regime, the total magnetization $\langle | m | \rangle $
essentially reaches its saturation value in the interval $H_{sc}-H_s
\propto 1/L^2$, and this is the same regime over which the
Curie-Weiss-like divergence of the susceptibility is rounded off.
For $(H_{sc}-H_s)L^2\gg 1$, however, $\langle | m | \rangle$ is
almost identical to its saturation value,
and the susceptibility converges towards the
(small) susceptibility of a bulk Ising system, independent of
size.
The maximum value of the
susceptibility scales like the system volume, $L^3$, 
as in a first order transition (Fig.~\ref{fig8}c), but unlike the
latter the shape of the susceptibility maximum does not converge
to a delta-function singularity (Fig.~\ref{fig8}b).

\item
A very special behavior is detected for the probability
distribution of the order parameter (Figs.~\ref{fig11} and
\ref{fig12}). On a scale of $|H_{sc}-H_{s}| \propto 1/L^2$ the
shape of this distribution changes from a single Gaussian peak
(for $H_s > H_{sc}$) to a double peak distribution (for
$H_s<H_{sc}$), and the inverse width $r(H_s)$ vanishes linearly
with $H_s$ as $H_{sc}$ is approached from above (see inset of
Fig.~\ref{fig11}a). This transition from single to double peak
shape happens via distribution $P_L(m) \propto \exp[-am^4]$, with
a coefficient $a$ that is {\em independent} of $L$ (Fig.~\ref{fig12}a).
This broadness of the distribution implies that in the limit $L
\rightarrow \infty$ at $H_s=H_{sc}$ fluctuations of the
magnetization occur which have a macroscopic, size-independent
amplitude. The standard statement of statistical thermodynamics,
that in the thermodynamic limit the relative magnitude of
fluctuations $(|\Delta m|/m_b)$ is negligibly small is not at all
true here. Again the behavior is completely different from a
standard first-order transition: at the latter, the system would
jump between $m=0$ and $m=-m_b$ and $+m_b$, whereas at the 
pathological second order transition found here the magnetization
can fluctuate over a finite fraction of the interval between 
$-m_b$ and $+m_b$, characterized by the distribution of Fig.~\ref{fig12}a. 
At a first order transition, we
would instead have three delta functions (at $m= \pm m_b$ and
$m=0$, respectively) as an order parameter distribution. However,
this particular behavior is easily accounted for by the
Landau-type theory of Sec.~II. In particular, for $H_s<H_{sc}$ the
simulations confirm the behavior $\ln P_L \propto L^{-2}m^2$
near the minimum at $m=0$ (Fig.~\ref{fig12}c).

\item In the regime for $H_s<H_{sc}$, when the system is large
enough so that $|m|$ is close to its saturation value $m_b$, the
variation of the interface distance $\ell_0$ counted from one of the
corners is found to go as $\ell _0 \propto (H_{sc}-H_s)^{-1}$, as
predicted by Parry et al. \cite{21}. When $\ell _0$ becomes
comparable to $L/2$, this divergence is rounded off, as expected
from the behavior of $P_L(m)$, since then a crossover from the
theory of Ref.~\cite{21} to the behavior described by our
Landau-like theory occurs. However, the detailed behavior in this
crossover regions is not yet fully understood.

\item We now discuss the extent to which similar behavior
can be expected to be found for real systems, such as 
the liquid-gas transition in a suitable
cavity, where half of the surface area has an energetic preference
for the liquid (such that ``incomplete wetting'' of the liquid at
the wall occurs) and the other half prefers the gas (i.e., an
``incomplete drying'' boundary condition). In this case one
neither has a precise symmetry between liquid and gas in the bulk,
nor can one expect a precise antisymmetry between the interactions
at the two types of walls. Therefore the phase transitions 
will be shifted somewhat away from the chemical potential
value at which phase coexistence can occur in the bulk. A similar
smooth interpolation between ``capillary condensation''-like
behavior and ``interface localization-delocalization'' transitions
has also been found for a model of a polymer blend in a thin film
geometry confined between parallel plates at which surface fields
act which do not have a particular symmetry \cite{78}. 
It is encouraging that wetting phenomena in systems 
where chemically distinct substrates meet find increasing theoretical \cite{25} and experimental
attention\cite{81}. An important constraint though is that for the fluids
one has to consider the grand-canonical ensemble where
the fluid  in the cavity can exchange particles with a reservoir,
and similarly for the binary fluid in the cavity also exchanges $A
\leftrightharpoons B$ or vice versa must be possible, due to a connection
with a suitable reservoir. If one considers a fluid in a cavity
with a fixed total number of fluid particles, or a binary mixture
in a cavity with fixed relative concentration, this transition of
Fig.~\ref{fig1} is completely suppressed: this situation would
correspond to an Ising system at constant total magnetization, and
hence by construction fluctuations of the uniform magnetization
then are impossible. In Fig.~\ref{fig1} then the configuration
with an interface present in the basal plane of the bi-pyramid is
enforced at all temperatures. An interesting aspect is also the
crossover between the transition studied here and the standard 
critical behavior near the bulk critical temperature (a crossover
from wetting to critical adsorption\cite{82}).
\end{enumerate}

\underline{Acknowledgments}: This work has greatly benefitted
from stimulating discussions with Andrew O. Parry, Amnon
Aharony and Siegfried Dietrich. One of us (A. M.) is grateful to the Deutsche
Forschungsgemeinschaft for support under grant No 436 BUL 113/130.

%\newpage


\begin{thebibliography}{99}
\bibitem{62} Note that in Ref. \cite{31} the factor $8/(3k_BT)$
was absorbed in the definition of $f_L(m)$.
\bibitem{66} The authors are greatly indebted to Prof. A. Aharony
for a helpful discussion of this issue.
\end{thebibliography}
\end{document}